\documentclass[aps,prl,twocolumn,nobibnotes,superscriptaddress,notitlepage]{revtex4-2}
\usepackage[hyperindex,breaklinks,colorlinks,urlcolor=JungleGreen,citecolor=JungleGreen,linkcolor=JungleGreen]{hyperref}
\usepackage{graphicx}
\usepackage{ulem}
\usepackage{textgreek}
\usepackage{natbib}
\usepackage{amsmath,blkarray}
\usepackage{bigstrut}
\usepackage{subfigure}
\usepackage{amsmath}
\usepackage{dcolumn}% Align table columns on decimal point
\usepackage{bm}% bold math
\usepackage[dvipsnames]{xcolor}
\def\ket#1{|#1\rangle}

\usepackage{comment}

\newcommand{\huPhys}{Department of Physics, Harvard University, Cambridge, Massachusetts 02138, USA}
\newcommand{\huSEAS}{John A. Paulson School of Engineering and Applied Sciences, Harvard University, Cambridge, Massachusetts 02138, USA}
\newcommand{\huChem}{Department of Chemistry and Chemical Biology, Harvard University, Cambridge, Massachusetts 02138, USA}
\newcommand{\umdPhys}{Department of Physics, University of Maryland, College Park, Maryland 20742, USA}
\newcommand{\umdECE}{Department of Electrical and Computer Engineering, University of Maryland, College Park, Maryland 20742, USA}
\newcommand{\umdQTC}{Quantum Technology Center, University of Maryland, College Park, Maryland 20742, USA}

\newcommand{\ICpostdoc}{Intelligence Community Postdoctoral Research Fellowship Program, University of Maryland, College Park, Maryland 20742, USA}

\newcommand{\tum}{Department of Chemistry, Technical University of Munich, Germany}

\newcommand{\NV}{NV\ }
\newcommand{\NVnospace}{NV}
\newcommand\isotope[2]{\textsuperscript{#2}#1}

\begin{document}

\title{Quantum Logic Enhanced Sensing in Solid-State Spin Ensembles}

\date{\today}

\author{Nithya Arunkumar}
\affiliation{\huPhys}
\affiliation{\huSEAS}
\affiliation{\umdQTC}

\author{Kevin S. Olsson}
\affiliation{\umdQTC}
\affiliation{\umdECE}
\affiliation{\ICpostdoc}

\author{Jner Tzern Oon}
\affiliation{\umdQTC}
\affiliation{\umdPhys}

\author{Connor Hart}
\affiliation{\umdQTC}
\affiliation{\umdECE}

\author{Dominik B. Bucher}
\affiliation{\huPhys}
\affiliation{\tum}

\author{David Glenn}
\affiliation{\huPhys}

\author{Mikhail D. Lukin}
\affiliation{\huPhys}

\author{Hongkun Park}
\affiliation{\huPhys}
\affiliation{\huChem}

\author{Donhee Ham }
\affiliation{\huSEAS}

\author{Ronald L. Walsworth}
\thanks{walsworth@umd.edu}
\affiliation{\huPhys}
\affiliation{\umdQTC}
\affiliation{\umdECE}
\affiliation{\umdPhys}

\date{\today}

\begin{abstract}

We demonstrate quantum logic enhanced sensitivity for a macroscopic ensemble of solid-state, hybrid two-qubit  sensors. We achieve a factor of 30 improvement in signal-to-noise ratio, translating to a sensitivity enhancement exceeding an order of magnitude. Using the electronic spins of nitrogen vacancy (\NVnospace) centers in diamond as sensors, we leverage the on-site nitrogen nuclear spins of the \NV centers as memory qubits, in combination with homogeneous bias and control fields, ensuring that all of the ${\sim}10^9$ two-qubit sensors are sufficiently identical to permit global control of the NV ensemble spin states.
We find quantum logic sensitivity enhancement for multiple measurement protocols with varying optimal sensing intervals, including XY8 dynamical decoupling and correlation spectroscopy, using a synthetic AC magnetic field. The results are independent of the nature of the target signal and broadly applicable to metrology using \NV centers and other solid-state ensembles. This work provides a benchmark for
macroscopic ensembles of quantum sensors that employ quantum logic or quantum error correction algorithms for enhanced sensitivity.

\end{abstract}

\maketitle

Quantum technologies for information processing, networking, and metrology increasingly employ hybrid architectures leveraging the advantageous properties of multiple physical qubit types. In platforms for quantum computing, this approach enables increased system sizes, while maintaining high fidelity and low cross-talk~\cite{singh_dual_neutral_atom, bruzewicz_trapped_ions, inlek_trapped_ions,arute_google_sc}. For timekeeping applications, ion clocks using two species, one `clock' ion with a stable transition and a second `logic' ion for preparation and readout, have demonstrated exceptional performance~\cite{wineland_clocks, rosenband_clock_2007, hume_clock_2007,rosenband_clock_2008}.

Analogous multi-qubit strategies are similarly compelling for quantum sensors. In particular, the use of additional memory qubits in support of sensing qubits presents an attractive avenue towards entanglement-enhanced performance~\cite{Barry2020, hopper_review}. Among leading quantum sensing platforms, defect centers in solid-state systems are naturally amenable to quantum logic assisted protocols since they commonly consist of multiple individually controllable spin degrees of freedom localized within or near a defect.

An early demonstration of entanglement-enhanced sensing was by Jiang \textit{et al.}~\cite{Jiang2009}, who performed repetitive, quantum non-demolition readout of a single-negatively charged nitrogen-vacancy (\NVnospace) center spin state coupled to nearby \isotope{C}{13} nuclear spins.
The technique circumvents the poor optical readout fidelity of \NV centers by mapping the \NV electronic spin state onto the longer-lived nuclear memory qubit. The nuclear spin state can then be interrogated repeatedly by re-entangling the NV electronic and \isotope{C}{13} nuclear spins between each optical readout.

Subsequent work with single \NV centers improved upon this protocol by using the nitrogen nuclear spin inherent to the \NV center as a memory qubit, instead of the \isotope{C}{13} nuclear spin~\cite{Neumann2010,Lovchinsky2016}. In contrast to \isotope{C}{13} nuclear spins, which are randomly distributed within the diamond lattice, the on-site nitrogen nuclear spin has a well-determined hyperfine coupling with the \NV electronic spin. This homogeneous coupling makes it particularly suitable for translation to ensembles of \NV centers, as reported here.

By spatially averaging over many \NV centers, ensemble measurements trade nanometer-scale spatial resolution for dramatically improved sensitivity~\cite{Barry2020}, which has enabled a wide range of applications across the physical and life sciences, including by nuclear magnetic resonance (NMR)~\cite{Glenn2018}, magnetic microscopy~\cite{Levine2019}, crystal stress and pressure spectroscopy~\cite{Marshall2022,Hsieh2019}, and thermometry~\cite{Toyli2013,Neumann2013}. However, in nearly all such experiments, only global control of the \NV ensemble is available. Therefore, the useful realization of quantum logic protocols with \NV ensembles demands each constituent multi-qubit sensor system be nearly identical, for both interactions within the \NV center and diamond lattice, as well as its response to external fields.

In this Letter, we demonstrate quantum logic enhanced (QLE) sensing with a macroscopic ensemble of two-qubit sensors, each consisting of an \NV electronic spin and an on-site \isotope{N}{15} nuclear spin. Interrogating a $($10$\times$10$\times$10$)\,$\textmu{m}$^3$ volume of diamond containing ${\sim}10^9$ of these two-qubit sensors, the QLE protocol increases the effective readout fidelity of each constituent NV, achieving a nominal $33\times$ improvement in signal-to-noise ratio (SNR) compared to measurements using only the \NV electronic spins.

We also show that the observed SNR improvement translates to enhancements in magnetic field sensitivity, which can exceed an order-of-magnitude. Importantly, the present approach is universally applicable (independent of the sensing protocol or target signal), and thus provides a benchmark for quantum sensing using quantum logic architectures, both for ensembles of NV centers and other platforms.

\begin{figure}[t!]
	\centering
	\includegraphics[width=3.4 in]{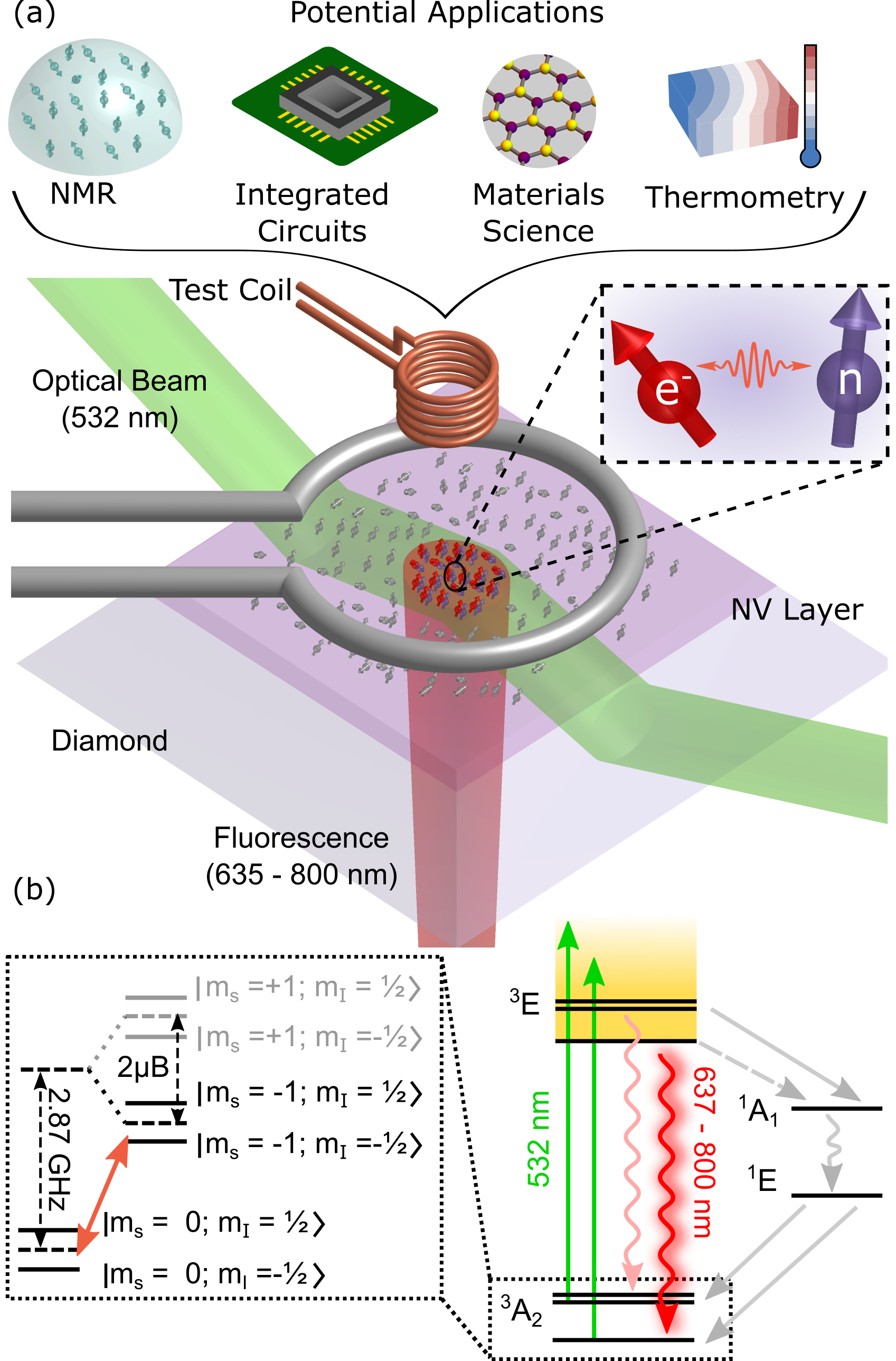}
	\caption{\textbf{\NV ensemble sensor integrated with quantum logic \cite{SI}.} (a) A 532\,nm optical beam illuminates the diamond chip with a spot diameter of $\sim$15\,\textmu{m}, using a total internal reflection configuration. A single-loop coil antenna above the $10\,$\textmu{m}-thick \NV layer drives both the microwave transition of the NV electronic spin states and the radio frequency transition of the \isotope{N}{15} nuclear spins. The magnified view depicts the multi-qubit \NV spin system. A multi-loop coil provides the synthetic AC magnetic field measured in this study, which can be substituted with a relevant target signal, such as those originating, e.g., from NMR, integrated circuits, materials science, and thermometry applications. (b) \NV energy levels allow optical initialization and readout of the electronic spin states. The expanded view of the ground state triplet energy levels shows splittings due to Zeeman and hyperfine interactions.
	}
	\label{fig:fig1}
\end{figure}

Figure~\ref{fig:fig1}(a) illustrates the experimental setup (refer to the supplement for further detail \cite{SI}), which operates at ambient laboratory conditions and probes a (2$\times$2$\times$0.5)\,mm$^3$ diamond chip containing an ensemble of \NV centers at a concentration of [NV]\,${\approx}\,2.3$\,ppm ([N]$\,{\approx}\,14$\,ppm)~\cite{Edmonds2021}. Each near-identical two-qubit system consists of the \NV electronic `sensor' spin ($S=1$) and the nuclear `memory' spin ($I=1/2$) of the associated \isotope{N}{15} nucleus. By applying a bias magnetic field along the \NV symmetry axis, the \NV ground state spin sublevels are non-degenerate (Fig.~\ref{fig:fig1}(b)), allowing them to be individually addressed using microwave fields. We employ the electronic spin states $m_s\!=\!0$ and $m_s\!=\!\text{-}1$ as an effective two-level system, with representative qubit states $\ket{\downarrow_e}$ and $\ket{\uparrow_e}$, respectively. The two nuclear spin states $m_I =  -\frac{1}{2}, \frac{1}{2}$ are represented by $\ket{\downarrow_n}$ and $\ket{\uparrow_e}$. The eigenstates of the composite electron-nuclear system are denoted by $\ket{ \{ \uparrow_e, \downarrow_e \}; \{ \uparrow_n, \downarrow_n \}}$.

\begin{figure}[t!]
	\centering
	\includegraphics[width=3.4 in]{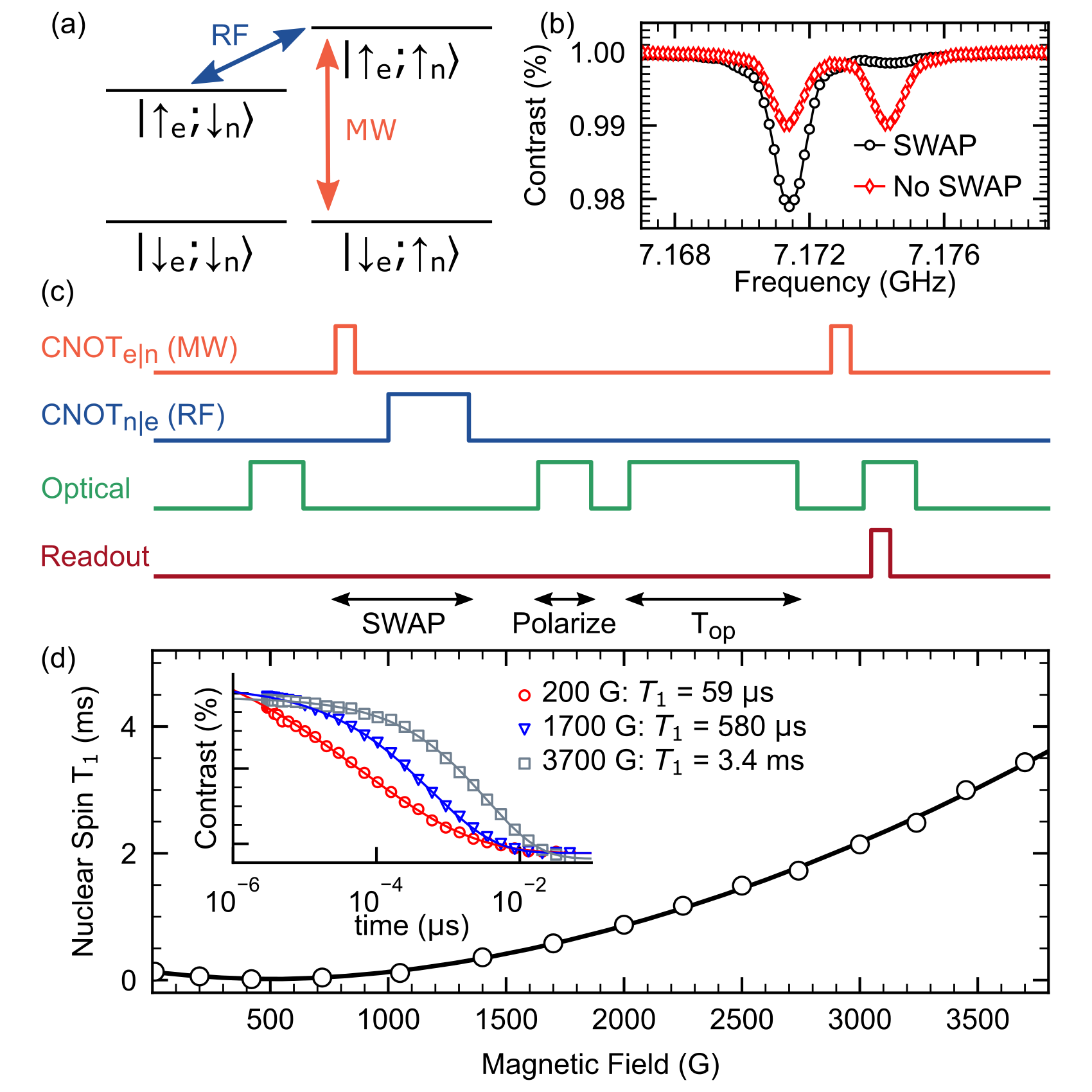}
	\caption{
		\textbf{Nuclear spin control with quantum logic. (a)} The energy levels of the two-qubit system are formed by the \NV electronic spin and its on-site \isotope{N}{15} nuclear spin. A selective MW pulse exchanges the spin population between $\ket{\downarrow_e; \uparrow_n}$ and $\ket{\uparrow_e; \uparrow_n}$, acting as a CNOT$_{e|n}$ gate. An RF pulse implements a CNOT$_{n|e}$ gate, exchanging the populations of $\ket{\uparrow_e; \uparrow_n}$ and $\ket{\uparrow_e; \downarrow_n}$. Applying the CNOT$_{e|n}$ and CNOT$_{n|e}$ gates in succession contitutes a SWAP operation. \textbf{(b)} Measured NV ensemble ODMR spectra with and without the SWAP operation.
		\textbf{(c)} Experimental pulse sequence used to measure the nuclear spin lifetime $T_1$ under optical illumination. \textbf{(d)} Measured $^{15}$N ensemble nuclear spin lifetime $T_1$ (white circles) at bias magnetic fields up to $3700$\,G, fit to a power law function.
		}
	\label{fig:fig2}
\end{figure}

\begin{figure}[t!]
	\centering
	\includegraphics[width=3.4in]{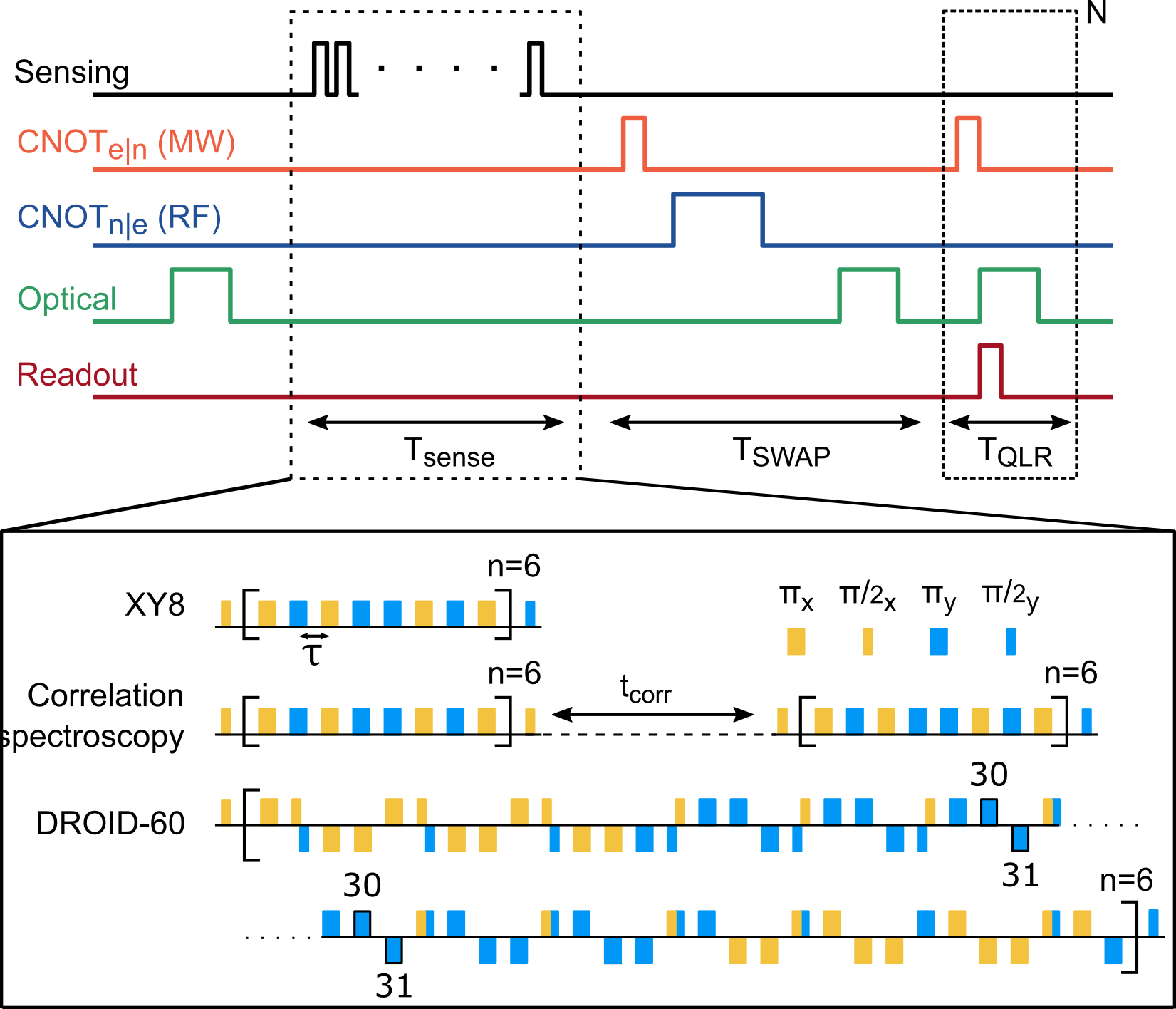}
	\caption{
		\textbf{Quantum logic enhanced (QLE) sensing protocol.} After a sensing sequence with a duration $T_\text{sense}$, the electronic spin state is encoded onto the nuclear spin for each NV in the ensemble, through the application of CNOT$_{e|n}$ and CNOT$_{n|e}$ gates. After a subsequent optical pulse polarizes the electronic spins, the readout sequence consisting of a CNOT$_{e|n}$ gate and a repolarization pulse is repeated $N$ times. The duration of each individual quantum logic readout sequence, $T_\text{QLR}$, is limited by the time required to repolarize the electronic spins. Example AC magnetic field sensing sequences are shown in the magnified view of $T_\text{sense}$.
		}
	\label{fig:fig3}
\end{figure}

We first demonstrate control of the \NV nuclear spin (memory qubit) using the \NV electronic spin (sensor qubit) via quantum logic protocol. To entangle the NV sensor and memory qubits, two controlled NOT (CNOT) gates are applied in succession, forming a SWAP operation. Fig.~\ref{fig:fig2}(a) and the top panel of Fig.~\ref{fig:fig3} (excluding the sensing sequence) illustrate this procedure, which begins with a selective microwave (MW) $\pi$ pulse that exchanges the spin populations of the states $\ket{\downarrow_e; \uparrow_n}$ and $\ket{\uparrow_e; \uparrow_n}$. This pulse acts as a CNOT operation on the electronic spin, conditioned on the nuclear spin state (CNOT$_{e|n}$). Next, a radio frequency (RF) $\pi$ pulse is applied, resulting in a CNOT$_{n|e}$ gate that exchanges the populations of $\ket{\uparrow_e; \uparrow_n}$ and $\ket{\uparrow_e; \downarrow_n}$. As a result, the electronic spin population is mapped to that of the nuclear spin, effectively encoding the information measured by the sensor spins onto the memory spins.

To evaluate the efficacy of the SWAP operation, we perform optically detected magnetic resonance (ODMR) measurements on the \NV ensemble. After applying an initial 532\,nm laser pulse to polarize the electronic spin states to $\ket{\downarrow_e}$, the ODMR spectra with and without the SWAP operation is measured, as shown in Figure~\ref{fig:fig2}(b). A clear transfer of polarization from the electronic sensor spins to the nuclear memory spins is observed when the SWAP operation is applied, with an estimated fidelity of $93\%$.

After encoding the sensor spin population onto the memory spins using the SWAP gate, the electronic spin states are reset using an optical polarization pulse. With successive $N$ applications of a CNOT$_{e|n}$ gate followed by an optical readout pulse, the information stored in the nuclear memory spins is then repeatedly mapped back onto the electronic spins and measured optically. This procedure provides many readouts within a duration limited by the nuclear spin lifetime $T_1$, thereby enhancing the overall readout fidelity.  The large number of NVs probed allows a high-precision ensemble average measurement of the sensor spin state with one execution of this repetitive readout protocol.

Flip-flop transitions between the electronic and nuclear spins present the dominant $T_1$ relaxation channel, and are increasingly suppressed at higher bias magnetic fields.
Using the pulse sequence shown in Fig.~\ref{fig:fig2}(c), we present measurements of $T_1$ in Fig.~\ref{fig:fig2}(d) for magnetic fields up to 4000\,G, generated by a feedback-stabilized electromagnet~\cite{Glenn2018}. At each magnetic field, the \NV fluorescence signal contrast is measured as a function of optical pulse duration ($T_\text{op}$), which is applied after the SWAP operation and electronic spin reset, as described above, and before a single application of the CNOT$_{e|n}$ gate. The contrast decay curves are fit to a stretched exponential function
%of the form
%$a*\exp{\left(-\left(\text{T}_{op}/T_1\right)^c\right)}+d$
to extract $T_1$, with examples at the magnetic fields 200\,G, 1700\,G, and 3700\,G shown in the inset of Fig.~\ref{fig:fig2}(d). The intensity of the laser pulse also affects the nuclear $T_1$ (see~\cite{SI}), and is kept fixed. The $T_1$ values are fit to a power law function, yielding a power law exponent of 1.8(2) in agreement with an expected quadratic dependence \cite{Neumann2010}.
The measurements presented in the remainder of this paper are performed at a bias magnetic field of 3700\,G, with a corresponding $^{15}$N nuclear spin $T_1$ of 3.44(12)\,ms.

\begin{figure}[t!]
	\centering
	\includegraphics[width=3.4in]{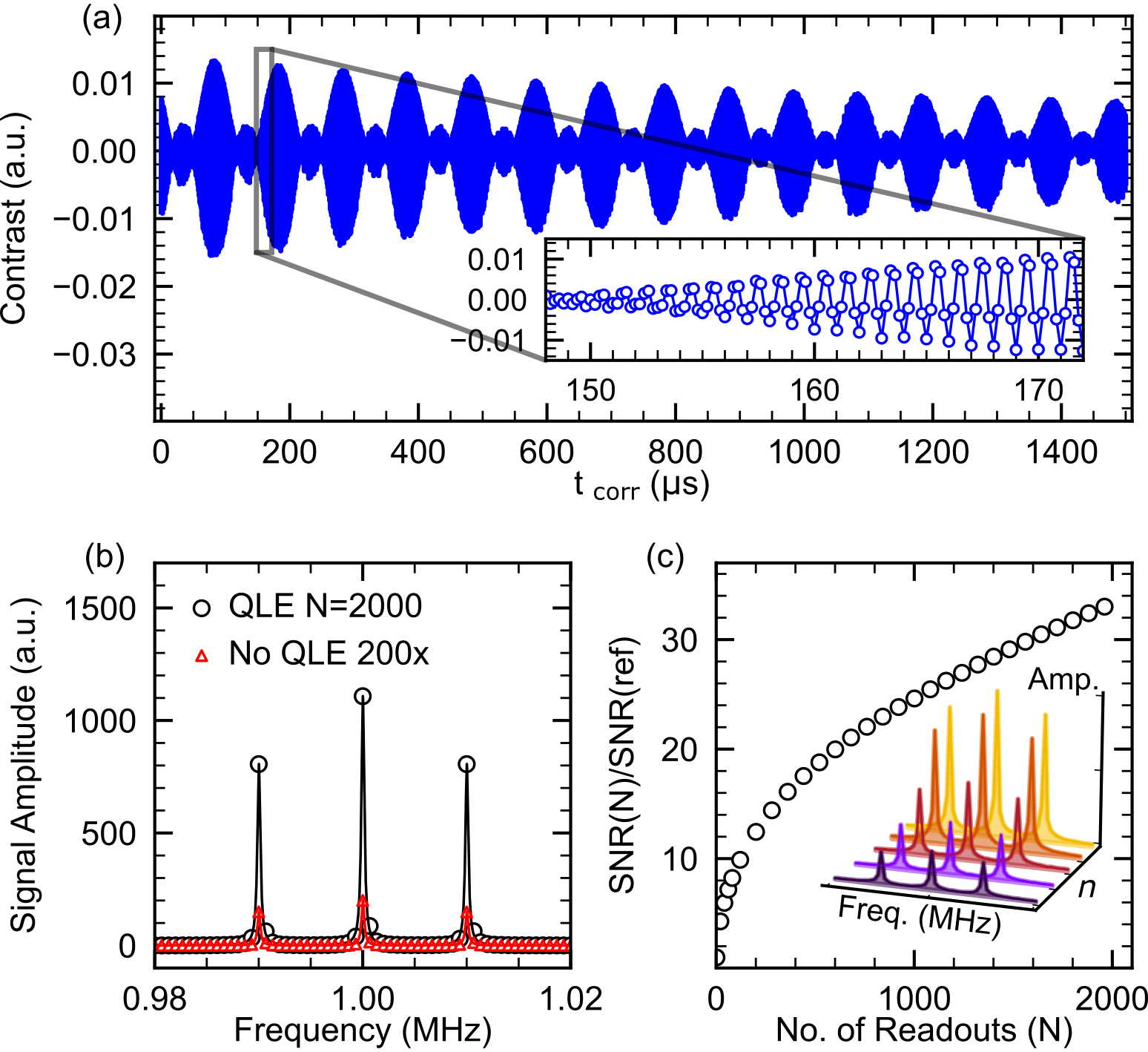}
	\caption{
		\textbf{Quantum logic enhanced (QLE) readout demonstrated with correlation spectroscopy.}
		\textbf{(a) }
		Example measured \NV correlation spectroscopy time trace of the three-tone test signal, without the QLE protocol.
		\textbf{(b)} Power spectra of the correlation spectroscopy signal with and without QLE protocol. The power spectrum in the absence of quantum logic is scaled $200\times$ for visibility.
		\textbf{(c)} Signal-to-noise ratio (SNR) of the measured test signal as a function of the number of quantum logic readout (QLR) cycles, $N$, compared to the reference SNR of conventional readout, i.e., a single readout sequence without the QLE protocol. Inset shows the decreasing signal amplitude from conventional readout (lightest) compared to the $n$-th QLR cycle, with $n$ = 1, 600, 1300, and 2000 (darkest).
		}
	\label{fig:fig4}
\end{figure}

We apply the above repetitive readout technique to demonstrate quantum logic enhanced (QLE) AC magnetic field sensing using an ensemble of NV centers. We first use correlation spectroscopy to measure a three-tone test signal at about 1\,MHz, generated by an RF synthesizer and communicated to the NV ensemble via a multi-loop coil (see Fig.~\ref{fig:fig1}, \cite{SI}).

Correlation spectroscopy is a popular \NV $T_1$-limited technique for AC magnetometry in which the time delay between two dynamical decoupling sequences ($T_\text{corr}$) is varied, see Fig.~\ref{fig:fig3}, with optical readout applied only after the second decoupling sequence~\cite{Laraoui2013,Bucher2019, Staudacher2015, Kong_corr, Kehayias2017}. Our correlation spectroscopy measurement employs two XY8-6 sequences~\cite{Gullion_xy8,Ryan_xy8}. As shown in Fig.~\ref{fig:fig3}, the sensing interval $T_\text{sense}$ is followed by a SWAP operation with duration $T_\text{SWAP}$ and $N$ quantum logic readout cycles (each with duration $T_\text{QLR}$).

Figure~\ref{fig:fig4}(a) depicts the NV ensemble fluorescence signal when measuring the test signal using correlation spectroscopy and conventional readout, which consists of a single optical readout pulse and omits other aspects of the QLE protocol. The corresponding power spectrum, with the three tones well-resolved, is shown in Fig.~\ref{fig:fig4}(b), scaled up by a factor of 200.

For the quantum logic enhanced (QLE) sensing protocol applied to measurements of the test signal, we determine a series of power spectra from the correlation time series acquired for each of the $N$ readout cycles. As apparent in the inset of Fig.~\ref{fig:fig4}(c), the power spectrum signal amplitudes, $A_n$, decay with increasing readout cycle index $n$ due to $^{15}$N nuclear spin $T_1$ relaxation. To optimize the signal-to-noise ratio SNR($N$) after $N$ readouts~\cite{Jiang2009}, the signal amplitude for the $n$-th readout is weighted by  $A_n/\sigma_n^2$, where $\sigma_n$ is the standard deviation of the noise at the $n$-th readout.
Fig.~\ref{fig:fig4}(b) compares the power spectrum of a weighted signal after $N=2000$ quantum logic readout (QLR) cycles to the reference signal obtained using conventional readout.

The resulting QLE SNR($N$) is given by
$\sqrt{\sum_{n=1}^{N} A_{n}^2/\sigma_{n}^2}.$
We normalize SNR($N$) by SNR(Ref), the SNR of the conventional \NV electronic spin readout (without quantum logic), measured under the same experimental conditions. The SNR enhancement realized with quantum logic is shown in Fig.~\ref{fig:fig4}(c). For example, with 2000 QLR cycles, we achieve a 33.3(9)$\times$ enhancement.

However, improvements in SNR do not necessarily translate into enhanced sensitivity, since SNR does not consider the impact on the measurement timescale (and hence bandwidth) of an extended readout interval. We first describe quantum logic sensitivity enhancement for AC magnetometry experiments using only a single dynamical decoupling sequence, and then address correlation spectroscopy experiments with varying $T_\text{sense}$. For non-correlation spectroscopy experiments, $A_n$ corresponds to the maximum signal amplitude as the magnitude of a single-tone, 1$\,$MHz test signal is varied.

Accounting for the overhead time associated with the SWAP operation ($T_\text{SWAP}=16.5\,$\textmu{s}) and each readout cycle ($T_\text{QLR}=3\,$\textmu{s}), the sensitivity enhancement obtained using the QLE protocol can be estimated from our measurements using,
\begin{equation}
\widetilde{\eta}_\text{QLE} \approx  \dfrac{\text{SNR}(N)}{ \text{SNR(Ref)}}\dfrac{\sqrt{T_\text{sense}+T_\text{QLR}}}{\sqrt{T_\text{sense}+T_\text{SWAP}+(N \times T_\text{QLR}})},
\label{eqn:equ3}
\end{equation}
where we have assumed the duration of a conventional readout is approximately $T_\text{QLR}$.

\begin{figure}[t]
	\centering
	\includegraphics[width=3.4in]{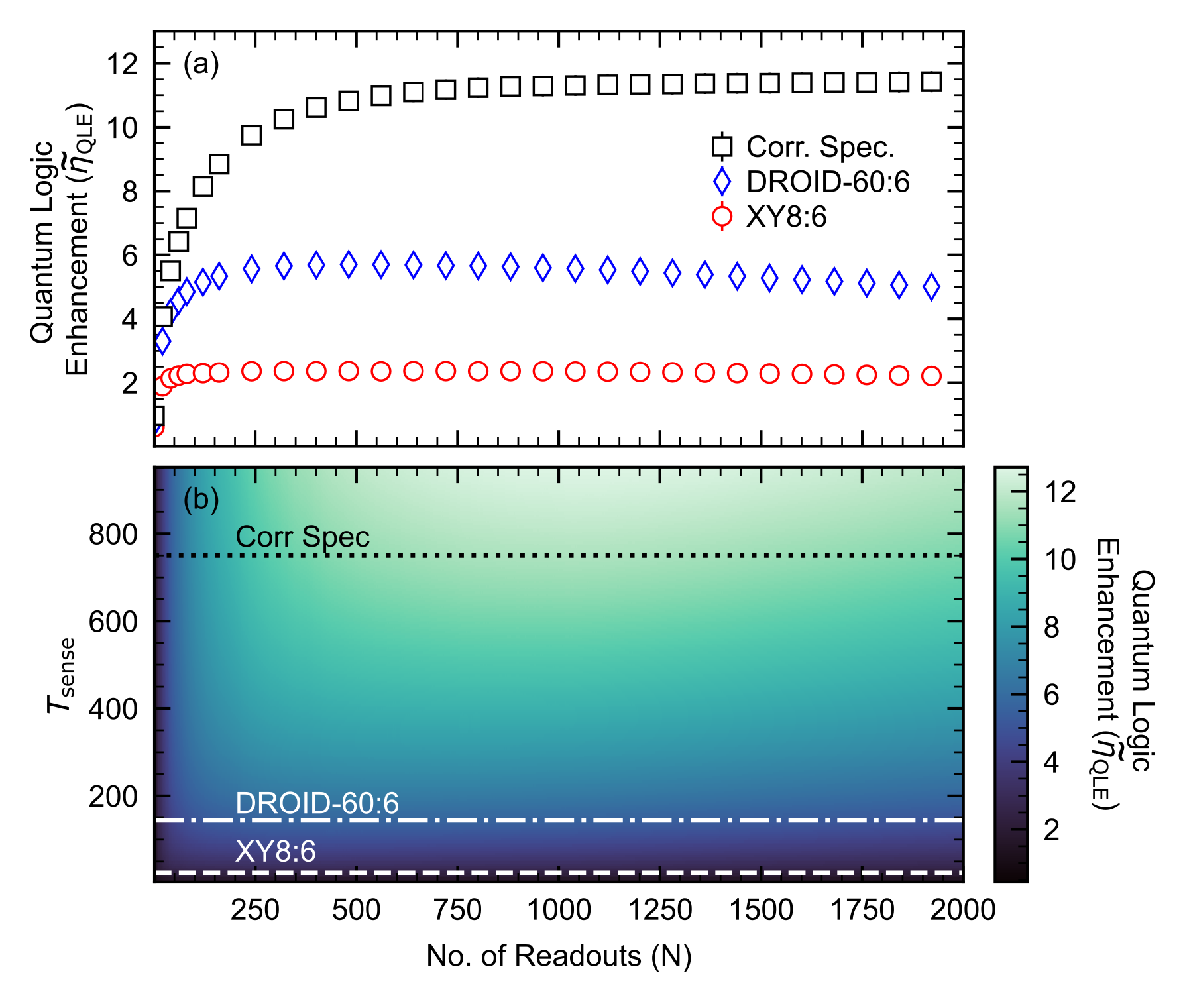}
	\caption{
		\textbf{Quantum logic enhanced (QLE) sensitivity.}
		\textbf{(a)} Experimentally determined QLE sensitivity factor ($\widetilde{\eta}_\text{QLE}$) as a function of the number $N$ of quantum logic readouts, for the three types of sensing sequences used in the present work.
		\textbf{(b)}
		Calculated QLE sensitivity factor ($\widetilde{\eta}_\text{QLE}$), given by color bar, as a function of the number of readouts $N$ and sensing duration $T_\text{sense}$. Calculations use Eq.~\ref{eqn:equ3} and the conditions of the present experiment. Dashed lines indicate the $T_\text{sense}$ values of the three sensing measurements reported in (a).
		}
	\label{fig:fig5}
\end{figure}

As shown in Fig.~\ref{fig:fig5}(a) for an XY8:6 (six repetitions of an XY8) dynamical decoupling sequence with an optimal $T_\text{sense}=24\,$\textmu{s}, the quantum logic protocol achieves up to $\widetilde{\eta}_\text{QLE}=2.4(3)$ for $N\approx 150$, compared to the same sensing sequence with conventional readout. For the diamond sample used here, the optimal $T_\text{sense}$ under XY8 decoupling is constrained by the \NV electronic spin coherence time $T_2\approx28\,$\textmu s, which, in turn, is limited by \NVnospace-\NV dipolar interactions \cite{SI}.

To surpass this interaction limit, we employ the DROID-60 decoupling sequence introduced in Refs.~\cite{Zhou2020,Choi2020} and thereby extend the optimal $T_\text{sense}$ to $144\,$\textmu{s}~\cite{SI}. As anticipated from Eq.~\eqref{eqn:equ3}, when the sensing duration becomes long compared to the additional overhead time for quantum logic readout, larger sensitivity enhancements are attained.
From our measurements we find up to $\widetilde{\eta}_\text{QLE}=5.6(3)$ for DROID-60:6 using quantum logic and $N\approx 400$, compared to conventional readout, as shown in Fig.~\ref{fig:fig5}(a).

For AC magnetometry using the correlation spectroscopy sequence described earlier, $T_\text{sense}$ varies with $T_\text{corr}$. To account for this dynamic $T_\text{sense}$ when acquiring correlation spectroscopy signals, we compare measurements using quantum logic with $N$ readouts to conventional measurements repeated $M$ times, where the total acquisition time for both experiments is equivalent. In Fig.~\ref{fig:fig5}(a), the number of conventional measurements $M$ used to calculate the reference sensitivity for correlation spectroscopy scales with $N$ to maintain: $M\,\times\,(T_\text{sense}+T_\text{QLR})\,=\,T_\text{sense}+T_\text{SWAP}+(N \times T_\text{QLR}$). For $T_\text{corr}$ ranging from $ 0\,\text{--}\,1.5\,$ms in our experiments, we find a quantum logic sensitivity enhancement of up to $\widetilde{\eta}_\text{QLE}=11.3(3)$ for $N \approx 1000$. More intuitively, if we use the average value of $T_\text{corr}$ (0.75\,ms) when calculating $\widetilde{\eta}_\text{QLE}$ via Eq.~\eqref{eqn:equ3}, we estimate a similar QLE sensitivity factor of $\widetilde{\eta}_\text{QLE} \approx 11$.

To highlight the versatility of the QLE technique, Fig.~\ref{fig:fig5}(b) provides estimates of quantum logic enhancement for a range of sensing durations T$_\text{sense}$, given our experimental conditions. Applying an arbitrarily large amount of readouts is not favorable. In \ref{fig:fig5}(b)  $\widetilde{\eta}_{\text{QLE}}$ is reduced as $N$ increases, most visible at $T_{\text{sense}} \sim 200 - 600 $ \textmu s for the ranges used in the plot. An improvement in sensitivity ($\widetilde{\eta}_\text{QLE}\!>\!1$) is readily achieved when $T_\text{sense}$ exceeds $T_\text{SWAP}$, applicable to a wide range of sensing sequences (and diamond materials) commonly used in \NVnospace-ensemble metrology and, in principle, for other solid-state spin systems.

In summary, we leveraged quantum logic using a macroscopic ensemble of solid-state, hybrid two-qubit sensors -- each consisting of an NV electronic spin and the on-site \isotope{N}{15} nuclear spin in diamond -- to realize a factor of 30 improvement in spin state readout SNR, which in turn enables significant improvement in AC magnetic field sensitivity. The observed sensitivity enhancements can exceed an order of magnitude under favorable conditions (i.e, sensing interval $\sim1$\,ms) using only global control of the \NV ensemble.

Over a range of nitrogen doping concentrations (0.5--20\,ppm), it is reasonable to tune the optimal sensing duration by decreasing the total nitrogen density in the material, which increases the \NV electronic coherence time ($T_2$) proportionally~\cite{Bauch2020}, without impacting the photon-shot-noise-limited sensitivity (assuming a constant N-\NV conversion efficiency). For example, exchanging the sample used here ([N$_\text{tot}$]$\,\approx$\,14\,ppm) for another with [N$_\text{tot}$]$\,\approx$\,0.8\,ppm in future work, we anticipate an 18-fold increase in the optimal XY8 sensing duration, sufficient to provide an order-of-magnitude improvement in quantum logic enhanced sensitivity (QLE) for the conditions otherwise used in the present experiments.

Furthermore, the current approach is agnostic to the target signal and, therefore, broadly applicable to sensing a variety of physical quantities. In the context of \NV ensemble magnetometry, the QLE protocol is naturally amenable to NMR spectroscopy given that both favor strong, uniform bias magnetic fields. Using a diamond with reduced [N] (and [\NVnospace]) to extend the \NV ensemble coherence time, the QLE protocol provides a path towards order-of-magnitude improvements in the sensitivity of micron-scale NV-NMR with high spectral resolution~\cite{Glenn2018}. Additionally, the improvements realized here are compatible with the growing collection of techniques for \NV-NMR sample hyperpolarization~\cite{Bucher2018,Arunkumar2021}.

Beyond magnetometry, NV-diamond dynamical decoupling protocols sensitive to crystal stress, pressure, and temperature have attained sensing durations of tens of microseconds or longer. For example, such sequences were recently employed in path-finding experiments for diamond-based dark matter searches~\cite{Marshall2022}. With further development, the key metric of $T_\text{sense} > T_\text{SWAP}$ may be realized for these alternative sensing applications, enabling quantum logic enhanced sensitivity.

Integrating additional quantum degrees of freedom, e.g., defects with couplings to multiple nuclear spins, is another promising direction for further progress in QLE sensing~\cite{Castelletto2020,Siyushev2014,Sajid2018,Udvarhelyi2017}. Similarly, solutions to address the random distribution of host lattice nuclear spins, such as manipulating the collective modes of a spin bath~\cite{Ruskuc2022}, may prove advantageous when exploring more advanced quantum logic or error correction algorithms~\cite{degen_review} for further sensing enhancements.

\begin{acknowledgements}
This work was supported by, or in part by, the U.S. Army Research Laboratory under Grant No.~W911NF1510548; the U.S. Army Research Office under Grant No.~W911NF1920181; the Moore Foundation; the DARPA DRINQS program under Grant No.~D18AC00033; and the University of Maryland Quantum Technology Center. D.B.B. was partially sup- ported by the German Research Foundation (BU 3257/ 1-1). K.S.O. acknowledges support through an appointment to the Intelligence Community Postdoctoral Research Fellowship Program at the University of Maryland, administered by Oak Ridge Institute for Science and Education through an interagency agreement between the U.S. Department of Energy and the Office of the Director of National Intelligence.
% N.A., D.B.B., M.J.T., and D.R.G. modified the NV-NMR spectrometer for the SABRE technique. N.A. and D.B.B. performed the experiments and N.A., K.S.O., J.O., C.H. analyzed the data. M.D.L, H.P., and R.L.W. conceived the application of NV diamond magnetometry to NMR detection at short length scales. R.L.W. supervised the project, including development of the overall experimental design. All authors discussed the results and participated in writing the manuscript.
\end{acknowledgements}

\bibliography{QLE_bibfile}

\begin{thebibliography}{39}%
\makeatletter
\providecommand \@ifxundefined [1]{%
 \@ifx{#1\undefined}
}%
\providecommand \@ifnum [1]{%
 \ifnum #1\expandafter \@firstoftwo
 \else \expandafter \@secondoftwo
 \fi
}%
\providecommand \@ifx [1]{%
 \ifx #1\expandafter \@firstoftwo
 \else \expandafter \@secondoftwo
 \fi
}%
\providecommand \natexlab [1]{#1}%
\providecommand \enquote  [1]{``#1''}%
\providecommand \bibnamefont  [1]{#1}%
\providecommand \bibfnamefont [1]{#1}%
\providecommand \citenamefont [1]{#1}%
\providecommand \href@noop [0]{\@secondoftwo}%
\providecommand \href [0]{\begingroup \@sanitize@url \@href}%
\providecommand \@href[1]{\@@startlink{#1}\@@href}%
\providecommand \@@href[1]{\endgroup#1\@@endlink}%
\providecommand \@sanitize@url [0]{\catcode `\\12\catcode `\$12\catcode
  `\&12\catcode `\#12\catcode `\^12\catcode `\_12\catcode `\%12\relax}%
\providecommand \@@startlink[1]{}%
\providecommand \@@endlink[0]{}%
\providecommand \url  [0]{\begingroup\@sanitize@url \@url }%
\providecommand \@url [1]{\endgroup\@href {#1}{\urlprefix }}%
\providecommand \urlprefix  [0]{URL }%
\providecommand \Eprint [0]{\href }%
\providecommand \doibase [0]{https://doi.org/}%
\providecommand \selectlanguage [0]{\@gobble}%
\providecommand \bibinfo  [0]{\@secondoftwo}%
\providecommand \bibfield  [0]{\@secondoftwo}%
\providecommand \translation [1]{[#1]}%
\providecommand \BibitemOpen [0]{}%
\providecommand \bibitemStop [0]{}%
\providecommand \bibitemNoStop [0]{.\EOS\space}%
\providecommand \EOS [0]{\spacefactor3000\relax}%
\providecommand \BibitemShut  [1]{\csname bibitem#1\endcsname}%
\let\auto@bib@innerbib\@empty
%</preamble>
\bibitem [{\citenamefont {Singh}\ \emph {et~al.}(2022)\citenamefont {Singh},
  \citenamefont {Anand}, \citenamefont {Pocklington}, \citenamefont {Kemp},\
  and\ \citenamefont {Bernien}}]{singh_dual_neutral_atom}%
  \BibitemOpen
  \bibfield  {author} {\bibinfo {author} {\bibfnamefont {K.}~\bibnamefont
  {Singh}}, \bibinfo {author} {\bibfnamefont {S.}~\bibnamefont {Anand}},
  \bibinfo {author} {\bibfnamefont {A.}~\bibnamefont {Pocklington}}, \bibinfo
  {author} {\bibfnamefont {J.~T.}\ \bibnamefont {Kemp}},\ and\ \bibinfo
  {author} {\bibfnamefont {H.}~\bibnamefont {Bernien}},\ }\bibfield  {title}
  {\bibinfo {title} {{Dual-Element, Two-Dimensional Atom Array with
  Continuous-Mode Operation}},\ }\href
  {https://doi.org/10.1103/PhysRevX.12.011040} {\bibfield  {journal} {\bibinfo
  {journal} {Phys. Rev. X}\ }\textbf {\bibinfo {volume} {12}},\ \bibinfo
  {pages} {11040} (\bibinfo {year} {2022})}\BibitemShut {NoStop}%
\bibitem [{\citenamefont {Bruzewicz}\ \emph {et~al.}(2019)\citenamefont
  {Bruzewicz}, \citenamefont {Chiaverini}, \citenamefont {McConnell},\ and\
  \citenamefont {Sage}}]{bruzewicz_trapped_ions}%
  \BibitemOpen
  \bibfield  {author} {\bibinfo {author} {\bibfnamefont {C.~D.}\ \bibnamefont
  {Bruzewicz}}, \bibinfo {author} {\bibfnamefont {J.}~\bibnamefont
  {Chiaverini}}, \bibinfo {author} {\bibfnamefont {R.}~\bibnamefont
  {McConnell}},\ and\ \bibinfo {author} {\bibfnamefont {J.~M.}\ \bibnamefont
  {Sage}},\ }\bibfield  {title} {\bibinfo {title} {{Trapped-ion quantum
  computing: Progress and challenges}},\ }\href
  {https://doi.org/10.1063/1.5088164} {\bibfield  {journal} {\bibinfo
  {journal} {Applied Physics Reviews}\ }\textbf {\bibinfo {volume} {6}},\
  \bibinfo {pages} {21314} (\bibinfo {year} {2019})}\BibitemShut {NoStop}%
\bibitem [{\citenamefont {Inlek}\ \emph {et~al.}(2017)\citenamefont {Inlek},
  \citenamefont {Crocker}, \citenamefont {Lichtman}, \citenamefont {Sosnova},\
  and\ \citenamefont {Monroe}}]{inlek_trapped_ions}%
  \BibitemOpen
  \bibfield  {author} {\bibinfo {author} {\bibfnamefont {I.~V.}\ \bibnamefont
  {Inlek}}, \bibinfo {author} {\bibfnamefont {C.}~\bibnamefont {Crocker}},
  \bibinfo {author} {\bibfnamefont {M.}~\bibnamefont {Lichtman}}, \bibinfo
  {author} {\bibfnamefont {K.}~\bibnamefont {Sosnova}},\ and\ \bibinfo {author}
  {\bibfnamefont {C.}~\bibnamefont {Monroe}},\ }\bibfield  {title} {\bibinfo
  {title} {{Multispecies Trapped-Ion Node for Quantum Networking}},\ }\href
  {https://doi.org/10.1103/PhysRevLett.118.250502} {\bibfield  {journal}
  {\bibinfo  {journal} {Phys. Rev. Lett.}\ }\textbf {\bibinfo {volume} {118}},\
  \bibinfo {pages} {250502} (\bibinfo {year} {2017})}\BibitemShut {NoStop}%
\bibitem [{\citenamefont {Arute}\ \emph {et~al.}(2019)\citenamefont {Arute},
  \citenamefont {Arya}, \citenamefont {Babbush}, \citenamefont {Bacon},
  \citenamefont {Bardin}, \citenamefont {Barends}, \citenamefont {Biswas},
  \citenamefont {Boixo}, \citenamefont {Brandao}, \citenamefont {Buell},
  \citenamefont {Burkett}, \citenamefont {Chen}, \citenamefont {Chen},
  \citenamefont {Chiaro}, \citenamefont {Collins} \emph
  {et~al.}}]{arute_google_sc}%
  \BibitemOpen
  \bibfield  {author} {\bibinfo {author} {\bibfnamefont {F.}~\bibnamefont
  {Arute}}, \bibinfo {author} {\bibfnamefont {K.}~\bibnamefont {Arya}},
  \bibinfo {author} {\bibfnamefont {R.}~\bibnamefont {Babbush}}, \bibinfo
  {author} {\bibfnamefont {D.}~\bibnamefont {Bacon}}, \bibinfo {author}
  {\bibfnamefont {J.~C.}\ \bibnamefont {Bardin}}, \bibinfo {author}
  {\bibfnamefont {R.}~\bibnamefont {Barends}}, \bibinfo {author} {\bibfnamefont
  {R.}~\bibnamefont {Biswas}}, \bibinfo {author} {\bibfnamefont
  {S.}~\bibnamefont {Boixo}}, \bibinfo {author} {\bibfnamefont {F.~G. S.~L.}\
  \bibnamefont {Brandao}}, \bibinfo {author} {\bibfnamefont {D.~A.}\
  \bibnamefont {Buell}}, \bibinfo {author} {\bibfnamefont {B.}~\bibnamefont
  {Burkett}}, \bibinfo {author} {\bibfnamefont {Y.}~\bibnamefont {Chen}},
  \bibinfo {author} {\bibfnamefont {Z.}~\bibnamefont {Chen}}, \bibinfo {author}
  {\bibfnamefont {B.}~\bibnamefont {Chiaro}}, \bibinfo {author} {\bibfnamefont
  {R.}~\bibnamefont {Collins}}, \emph {et~al.},\ }\bibfield  {title} {\bibinfo
  {title} {{Quantum supremacy using a programmable superconducting
  processor}},\ }\href {https://doi.org/10.1038/s41586-019-1666-5} {\bibfield
  {journal} {\bibinfo  {journal} {Nature}\ }\textbf {\bibinfo {volume} {574}},\
  \bibinfo {pages} {505} (\bibinfo {year} {2019})}\BibitemShut {NoStop}%
\bibitem [{\citenamefont {Schmidt}\ \emph {et~al.}(2005)\citenamefont
  {Schmidt}, \citenamefont {Rosenband}, \citenamefont {Langer}, \citenamefont
  {Itano}, \citenamefont {Bergquist},\ and\ \citenamefont
  {Wineland}}]{wineland_clocks}%
  \BibitemOpen
  \bibfield  {author} {\bibinfo {author} {\bibfnamefont {P.~O.}\ \bibnamefont
  {Schmidt}}, \bibinfo {author} {\bibfnamefont {T.}~\bibnamefont {Rosenband}},
  \bibinfo {author} {\bibfnamefont {C.}~\bibnamefont {Langer}}, \bibinfo
  {author} {\bibfnamefont {W.~M.}\ \bibnamefont {Itano}}, \bibinfo {author}
  {\bibfnamefont {J.~C.}\ \bibnamefont {Bergquist}},\ and\ \bibinfo {author}
  {\bibfnamefont {D.~J.}\ \bibnamefont {Wineland}},\ }\bibfield  {title}
  {\bibinfo {title} {{Spectroscopy Using Quantum Logic}},\ }\href
  {https://doi.org/10.1126/science.1114375} {\bibfield  {journal} {\bibinfo
  {journal} {Science}\ }\textbf {\bibinfo {volume} {309}},\ \bibinfo {pages}
  {749} (\bibinfo {year} {2005})}\BibitemShut {NoStop}%
\bibitem [{\citenamefont {Rosenband}\ \emph {et~al.}(2007)\citenamefont
  {Rosenband}, \citenamefont {Schmidt}, \citenamefont {Hume}, \citenamefont
  {Itano}, \citenamefont {Fortier}, \citenamefont {Stalnaker}, \citenamefont
  {Kim}, \citenamefont {Diddams}, \citenamefont {Koelemeij}, \citenamefont
  {Bergquist},\ and\ \citenamefont {Wineland}}]{rosenband_clock_2007}%
  \BibitemOpen
  \bibfield  {author} {\bibinfo {author} {\bibfnamefont {T.}~\bibnamefont
  {Rosenband}}, \bibinfo {author} {\bibfnamefont {P.~O.}\ \bibnamefont
  {Schmidt}}, \bibinfo {author} {\bibfnamefont {D.~B.}\ \bibnamefont {Hume}},
  \bibinfo {author} {\bibfnamefont {W.~M.}\ \bibnamefont {Itano}}, \bibinfo
  {author} {\bibfnamefont {T.~M.}\ \bibnamefont {Fortier}}, \bibinfo {author}
  {\bibfnamefont {J.~E.}\ \bibnamefont {Stalnaker}}, \bibinfo {author}
  {\bibfnamefont {K.}~\bibnamefont {Kim}}, \bibinfo {author} {\bibfnamefont
  {S.~A.}\ \bibnamefont {Diddams}}, \bibinfo {author} {\bibfnamefont
  {J.~C.~J.}\ \bibnamefont {Koelemeij}}, \bibinfo {author} {\bibfnamefont
  {J.~C.}\ \bibnamefont {Bergquist}},\ and\ \bibinfo {author} {\bibfnamefont
  {D.~J.}\ \bibnamefont {Wineland}},\ }\bibfield  {title} {\bibinfo {title}
  {Observation of the $^{1}$s$_{0}\ensuremath{\rightarrow}^{3}$p$_{0}$ clock
  transition in $^{27}\mathrm{Al}^{+}$},\ }\href
  {https://doi.org/10.1103/PhysRevLett.98.220801} {\bibfield  {journal}
  {\bibinfo  {journal} {Phys. Rev. Lett.}\ }\textbf {\bibinfo {volume} {98}},\
  \bibinfo {pages} {220801} (\bibinfo {year} {2007})}\BibitemShut {NoStop}%
\bibitem [{\citenamefont {Hume}\ \emph {et~al.}(2007)\citenamefont {Hume},
  \citenamefont {Rosenband},\ and\ \citenamefont {Wineland}}]{hume_clock_2007}%
  \BibitemOpen
  \bibfield  {author} {\bibinfo {author} {\bibfnamefont {D.~B.}\ \bibnamefont
  {Hume}}, \bibinfo {author} {\bibfnamefont {T.}~\bibnamefont {Rosenband}},\
  and\ \bibinfo {author} {\bibfnamefont {D.~J.}\ \bibnamefont {Wineland}},\
  }\bibfield  {title} {\bibinfo {title} {{High-Fidelity Adaptive Qubit
  Detection through Repetitive Quantum Nondemolition Measurements}},\ }\href
  {https://doi.org/10.1103/PhysRevLett.99.120502} {\bibfield  {journal}
  {\bibinfo  {journal} {Phys. Rev. Lett.}\ }\textbf {\bibinfo {volume} {99}},\
  \bibinfo {pages} {120502} (\bibinfo {year} {2007})}\BibitemShut {NoStop}%
\bibitem [{\citenamefont {Rosenband}\ \emph {et~al.}(2008)\citenamefont
  {Rosenband}, \citenamefont {Hume}, \citenamefont {Schmidt}, \citenamefont
  {Chou}, \citenamefont {Brusch}, \citenamefont {Lorini}, \citenamefont
  {Oskay}, \citenamefont {Drullinger}, \citenamefont {Fortier}, \citenamefont
  {Stalnaker}, \citenamefont {Diddams}, \citenamefont {Swann}, \citenamefont
  {Newbury}, \citenamefont {Itano}, \citenamefont {Wineland},\ and\
  \citenamefont {Bergquist}}]{rosenband_clock_2008}%
  \BibitemOpen
  \bibfield  {author} {\bibinfo {author} {\bibfnamefont {T.}~\bibnamefont
  {Rosenband}}, \bibinfo {author} {\bibfnamefont {D.~B.}\ \bibnamefont {Hume}},
  \bibinfo {author} {\bibfnamefont {P.~O.}\ \bibnamefont {Schmidt}}, \bibinfo
  {author} {\bibfnamefont {C.~W.}\ \bibnamefont {Chou}}, \bibinfo {author}
  {\bibfnamefont {A.}~\bibnamefont {Brusch}}, \bibinfo {author} {\bibfnamefont
  {L.}~\bibnamefont {Lorini}}, \bibinfo {author} {\bibfnamefont {W.~H.}\
  \bibnamefont {Oskay}}, \bibinfo {author} {\bibfnamefont {R.~E.}\ \bibnamefont
  {Drullinger}}, \bibinfo {author} {\bibfnamefont {T.~M.}\ \bibnamefont
  {Fortier}}, \bibinfo {author} {\bibfnamefont {J.~E.}\ \bibnamefont
  {Stalnaker}}, \bibinfo {author} {\bibfnamefont {S.~A.}\ \bibnamefont
  {Diddams}}, \bibinfo {author} {\bibfnamefont {W.~C.}\ \bibnamefont {Swann}},
  \bibinfo {author} {\bibfnamefont {N.~R.}\ \bibnamefont {Newbury}}, \bibinfo
  {author} {\bibfnamefont {W.~M.}\ \bibnamefont {Itano}}, \bibinfo {author}
  {\bibfnamefont {D.~J.}\ \bibnamefont {Wineland}},\ and\ \bibinfo {author}
  {\bibfnamefont {J.~C.}\ \bibnamefont {Bergquist}},\ }\bibfield  {title}
  {\bibinfo {title} {{Frequency Ratio of Al\textsuperscript{+} and
  Hg\textsuperscript{+} Single-Ion Optical Clocks; Metrology at the 17th
  Decimal Place}},\ }\href {https://doi.org/10.1126/science.1154622} {\bibfield
   {journal} {\bibinfo  {journal} {Science}\ }\textbf {\bibinfo {volume}
  {319}},\ \bibinfo {pages} {1808} (\bibinfo {year} {2008})}\BibitemShut
  {NoStop}%
\bibitem [{\citenamefont {Barry}\ \emph {et~al.}(2020)\citenamefont {Barry},
  \citenamefont {Schloss}, \citenamefont {Bauch}, \citenamefont {Turner},
  \citenamefont {Hart}, \citenamefont {Pham},\ and\ \citenamefont
  {Walsworth}}]{Barry2020}%
  \BibitemOpen
  \bibfield  {author} {\bibinfo {author} {\bibfnamefont {J.~F.}\ \bibnamefont
  {Barry}}, \bibinfo {author} {\bibfnamefont {J.~M.}\ \bibnamefont {Schloss}},
  \bibinfo {author} {\bibfnamefont {E.}~\bibnamefont {Bauch}}, \bibinfo
  {author} {\bibfnamefont {M.~J.}\ \bibnamefont {Turner}}, \bibinfo {author}
  {\bibfnamefont {C.~A.}\ \bibnamefont {Hart}}, \bibinfo {author}
  {\bibfnamefont {L.~M.}\ \bibnamefont {Pham}},\ and\ \bibinfo {author}
  {\bibfnamefont {R.~L.}\ \bibnamefont {Walsworth}},\ }\bibfield  {title}
  {\bibinfo {title} {Sensitivity optimization for nv-diamond magnetometry},\
  }\href {https://link.aps.org/doi/10.1103/RevModPhys.92.015004} {\bibfield
  {journal} {\bibinfo  {journal} {Rev. Mod. Phys.}\ }\textbf {\bibinfo {volume}
  {92}},\ \bibinfo {pages} {015004} (\bibinfo {year} {2020})}\BibitemShut
  {NoStop}%
\bibitem [{\citenamefont {Hopper}\ \emph {et~al.}(2018)\citenamefont {Hopper},
  \citenamefont {Shulevitz},\ and\ \citenamefont {Bassett}}]{hopper_review}%
  \BibitemOpen
  \bibfield  {author} {\bibinfo {author} {\bibfnamefont {D.~A.}\ \bibnamefont
  {Hopper}}, \bibinfo {author} {\bibfnamefont {H.~J.}\ \bibnamefont
  {Shulevitz}},\ and\ \bibinfo {author} {\bibfnamefont {L.~C.}\ \bibnamefont
  {Bassett}},\ }\bibfield  {title} {\bibinfo {title} {Spin readout techniques
  of the nitrogen-vacancy center in diamond},\ }\href
  {https://www.mdpi.com/2072-666X/9/9/437} {\bibfield  {journal} {\bibinfo
  {journal} {Micromachines}\ }\textbf {\bibinfo {volume} {9}} (\bibinfo {year}
  {2018})}\BibitemShut {NoStop}%
\bibitem [{\citenamefont {Jiang}\ \emph {et~al.}(2009)\citenamefont {Jiang},
  \citenamefont {Hodges}, \citenamefont {Maze}, \citenamefont {Maurer},
  \citenamefont {Taylor}, \citenamefont {Cory}, \citenamefont {Hemmer},
  \citenamefont {Walsworth}, \citenamefont {Yacoby}, \citenamefont {Zibrov},\
  and\ \citenamefont {Lukin}}]{Jiang2009}%
  \BibitemOpen
  \bibfield  {author} {\bibinfo {author} {\bibfnamefont {L.}~\bibnamefont
  {Jiang}}, \bibinfo {author} {\bibfnamefont {J.~S.}\ \bibnamefont {Hodges}},
  \bibinfo {author} {\bibfnamefont {J.~R.}\ \bibnamefont {Maze}}, \bibinfo
  {author} {\bibfnamefont {P.}~\bibnamefont {Maurer}}, \bibinfo {author}
  {\bibfnamefont {J.~M.}\ \bibnamefont {Taylor}}, \bibinfo {author}
  {\bibfnamefont {D.~G.}\ \bibnamefont {Cory}}, \bibinfo {author}
  {\bibfnamefont {P.~R.}\ \bibnamefont {Hemmer}}, \bibinfo {author}
  {\bibfnamefont {R.~L.}\ \bibnamefont {Walsworth}}, \bibinfo {author}
  {\bibfnamefont {A.}~\bibnamefont {Yacoby}}, \bibinfo {author} {\bibfnamefont
  {A.~S.}\ \bibnamefont {Zibrov}},\ and\ \bibinfo {author} {\bibfnamefont
  {M.~D.}\ \bibnamefont {Lukin}},\ }\bibfield  {title} {\bibinfo {title}
  {Repetitive readout of a single electronic spin via quantum logic with
  nuclear spin ancillae},\ }\href {https://doi.org/10.1126/science.1176496}
  {\bibfield  {journal} {\bibinfo  {journal} {Science}\ }\textbf {\bibinfo
  {volume} {326}},\ \bibinfo {pages} {267} (\bibinfo {year}
  {2009})}\BibitemShut {NoStop}%
\bibitem [{\citenamefont {Neumann}\ \emph {et~al.}(2010)\citenamefont
  {Neumann}, \citenamefont {Beck}, \citenamefont {Steiner}, \citenamefont
  {Rempp}, \citenamefont {Fedder}, \citenamefont {Hemmer}, \citenamefont
  {Wrachtrup},\ and\ \citenamefont {Jelezko}}]{Neumann2010}%
  \BibitemOpen
  \bibfield  {author} {\bibinfo {author} {\bibfnamefont {P.}~\bibnamefont
  {Neumann}}, \bibinfo {author} {\bibfnamefont {J.}~\bibnamefont {Beck}},
  \bibinfo {author} {\bibfnamefont {M.}~\bibnamefont {Steiner}}, \bibinfo
  {author} {\bibfnamefont {F.}~\bibnamefont {Rempp}}, \bibinfo {author}
  {\bibfnamefont {H.}~\bibnamefont {Fedder}}, \bibinfo {author} {\bibfnamefont
  {P.~R.}\ \bibnamefont {Hemmer}}, \bibinfo {author} {\bibfnamefont
  {J.}~\bibnamefont {Wrachtrup}},\ and\ \bibinfo {author} {\bibfnamefont
  {F.}~\bibnamefont {Jelezko}},\ }\bibfield  {title} {\bibinfo {title}
  {Single-shot readout of a single nuclear spin},\ }\href
  {https://doi.org/10.1126/science.1189075} {\bibfield  {journal} {\bibinfo
  {journal} {Science}\ }\textbf {\bibinfo {volume} {329}},\ \bibinfo {pages}
  {542} (\bibinfo {year} {2010})}\BibitemShut {NoStop}%
\bibitem [{\citenamefont {Lovchinsky}\ \emph {et~al.}(2016)\citenamefont
  {Lovchinsky}, \citenamefont {Sushkov}, \citenamefont {Urbach}, \citenamefont
  {de~Leon}, \citenamefont {Choi}, \citenamefont {De~Greve}, \citenamefont
  {Evans}, \citenamefont {Gertner}, \citenamefont {Bersin}, \citenamefont
  {M{\"u}ller}, \citenamefont {McGuinness}, \citenamefont {Jelezko},
  \citenamefont {Walsworth}, \citenamefont {Park},\ and\ \citenamefont
  {Lukin}}]{Lovchinsky2016}%
  \BibitemOpen
  \bibfield  {author} {\bibinfo {author} {\bibfnamefont {I.}~\bibnamefont
  {Lovchinsky}}, \bibinfo {author} {\bibfnamefont {A.~O.}\ \bibnamefont
  {Sushkov}}, \bibinfo {author} {\bibfnamefont {E.}~\bibnamefont {Urbach}},
  \bibinfo {author} {\bibfnamefont {N.~P.}\ \bibnamefont {de~Leon}}, \bibinfo
  {author} {\bibfnamefont {S.}~\bibnamefont {Choi}}, \bibinfo {author}
  {\bibfnamefont {K.}~\bibnamefont {De~Greve}}, \bibinfo {author}
  {\bibfnamefont {R.}~\bibnamefont {Evans}}, \bibinfo {author} {\bibfnamefont
  {R.}~\bibnamefont {Gertner}}, \bibinfo {author} {\bibfnamefont
  {E.}~\bibnamefont {Bersin}}, \bibinfo {author} {\bibfnamefont
  {C.}~\bibnamefont {M{\"u}ller}}, \bibinfo {author} {\bibfnamefont
  {L.}~\bibnamefont {McGuinness}}, \bibinfo {author} {\bibfnamefont
  {F.}~\bibnamefont {Jelezko}}, \bibinfo {author} {\bibfnamefont {R.~L.}\
  \bibnamefont {Walsworth}}, \bibinfo {author} {\bibfnamefont {H.}~\bibnamefont
  {Park}},\ and\ \bibinfo {author} {\bibfnamefont {M.~D.}\ \bibnamefont
  {Lukin}},\ }\bibfield  {title} {\bibinfo {title} {Nuclear magnetic resonance
  detection and spectroscopy of single proteins using quantum logic},\ }\href
  {https://science.sciencemag.org/content/351/6275/836} {\bibfield  {journal}
  {\bibinfo  {journal} {Science}\ }\textbf {\bibinfo {volume} {351}},\ \bibinfo
  {pages} {836} (\bibinfo {year} {2016})}\BibitemShut {NoStop}%
\bibitem [{\citenamefont {Glenn}\ \emph {et~al.}(2018)\citenamefont {Glenn},
  \citenamefont {Bucher}, \citenamefont {Lee}, \citenamefont {Lukin},
  \citenamefont {Park},\ and\ \citenamefont {Walsworth}}]{Glenn2018}%
  \BibitemOpen
  \bibfield  {author} {\bibinfo {author} {\bibfnamefont {D.~R.}\ \bibnamefont
  {Glenn}}, \bibinfo {author} {\bibfnamefont {D.~B.}\ \bibnamefont {Bucher}},
  \bibinfo {author} {\bibfnamefont {J.}~\bibnamefont {Lee}}, \bibinfo {author}
  {\bibfnamefont {M.~D.}\ \bibnamefont {Lukin}}, \bibinfo {author}
  {\bibfnamefont {H.}~\bibnamefont {Park}},\ and\ \bibinfo {author}
  {\bibfnamefont {R.~L.}\ \bibnamefont {Walsworth}},\ }\bibfield  {title}
  {\bibinfo {title} {High-resolution magnetic resonance spectroscopy using a
  solid-state spin sensor},\ }\href {https://doi.org/10.1038/nature25781}
  {\bibfield  {journal} {\bibinfo  {journal} {Nature}\ }\textbf {\bibinfo
  {volume} {555}},\ \bibinfo {pages} {351} (\bibinfo {year}
  {2018})}\BibitemShut {NoStop}%
\bibitem [{\citenamefont {Levine}\ \emph {et~al.}(2019)\citenamefont {Levine},
  \citenamefont {Turner}, \citenamefont {Kehayias}, \citenamefont {Hart},
  \citenamefont {Langellier}, \citenamefont {Trubko}, \citenamefont {Glenn},
  \citenamefont {Fu},\ and\ \citenamefont {Walsworth}}]{Levine2019}%
  \BibitemOpen
  \bibfield  {author} {\bibinfo {author} {\bibfnamefont {E.~V.}\ \bibnamefont
  {Levine}}, \bibinfo {author} {\bibfnamefont {M.~J.}\ \bibnamefont {Turner}},
  \bibinfo {author} {\bibfnamefont {P.}~\bibnamefont {Kehayias}}, \bibinfo
  {author} {\bibfnamefont {C.~A.}\ \bibnamefont {Hart}}, \bibinfo {author}
  {\bibfnamefont {N.}~\bibnamefont {Langellier}}, \bibinfo {author}
  {\bibfnamefont {R.}~\bibnamefont {Trubko}}, \bibinfo {author} {\bibfnamefont
  {D.~R.}\ \bibnamefont {Glenn}}, \bibinfo {author} {\bibfnamefont {R.~R.}\
  \bibnamefont {Fu}},\ and\ \bibinfo {author} {\bibfnamefont {R.~L.}\
  \bibnamefont {Walsworth}},\ }\bibfield  {title} {\bibinfo {title} {Principles
  and techniques of the quantum diamond microscope},\ }\href
  {https://doi.org/https://doi.org/10.1515/nanoph-2019-0209} {\bibfield
  {journal} {\bibinfo  {journal} {Nanophotonics}\ }\textbf {\bibinfo {volume}
  {8}},\ \bibinfo {pages} {1945 } (\bibinfo {year} {2019})}\BibitemShut
  {NoStop}%
\bibitem [{\citenamefont {Marshall}\ \emph {et~al.}(2022)\citenamefont
  {Marshall}, \citenamefont {Ebadi}, \citenamefont {Hart}, \citenamefont
  {Turner}, \citenamefont {Ku}, \citenamefont {Phillips},\ and\ \citenamefont
  {Walsworth}}]{Marshall2022}%
  \BibitemOpen
  \bibfield  {author} {\bibinfo {author} {\bibfnamefont {M.~C.}\ \bibnamefont
  {Marshall}}, \bibinfo {author} {\bibfnamefont {R.}~\bibnamefont {Ebadi}},
  \bibinfo {author} {\bibfnamefont {C.}~\bibnamefont {Hart}}, \bibinfo {author}
  {\bibfnamefont {M.~J.}\ \bibnamefont {Turner}}, \bibinfo {author}
  {\bibfnamefont {M.~J.}\ \bibnamefont {Ku}}, \bibinfo {author} {\bibfnamefont
  {D.~F.}\ \bibnamefont {Phillips}},\ and\ \bibinfo {author} {\bibfnamefont
  {R.~L.}\ \bibnamefont {Walsworth}},\ }\bibfield  {title} {\bibinfo {title}
  {High-precision mapping of diamond crystal strain using quantum
  interferometry},\ }\href {https://doi.org/10.1103/PhysRevApplied.17.024041}
  {\bibfield  {journal} {\bibinfo  {journal} {Phys. Rev. Applied}\ }\textbf
  {\bibinfo {volume} {17}},\ \bibinfo {pages} {024041} (\bibinfo {year}
  {2022})}\BibitemShut {NoStop}%
\bibitem [{\citenamefont {Hsieh}\ \emph {et~al.}(2019)\citenamefont {Hsieh},
  \citenamefont {Bhattacharyya}, \citenamefont {Zu}, \citenamefont {Mittiga},
  \citenamefont {Smart}, \citenamefont {Machado}, \citenamefont {Kobrin},
  \citenamefont {H{\"{o}}hn}, \citenamefont {Rui}, \citenamefont {Kamrani},
  \citenamefont {Chatterjee}, \citenamefont {Choi}, \citenamefont {Zaletel},
  \citenamefont {Struzhkin}, \citenamefont {Moore}, \citenamefont {Levitas},
  \citenamefont {Jeanloz},\ and\ \citenamefont {Yao}}]{Hsieh2019}%
  \BibitemOpen
  \bibfield  {author} {\bibinfo {author} {\bibfnamefont {S.}~\bibnamefont
  {Hsieh}}, \bibinfo {author} {\bibfnamefont {P.}~\bibnamefont
  {Bhattacharyya}}, \bibinfo {author} {\bibfnamefont {C.}~\bibnamefont {Zu}},
  \bibinfo {author} {\bibfnamefont {T.}~\bibnamefont {Mittiga}}, \bibinfo
  {author} {\bibfnamefont {T.~J.}\ \bibnamefont {Smart}}, \bibinfo {author}
  {\bibfnamefont {F.}~\bibnamefont {Machado}}, \bibinfo {author} {\bibfnamefont
  {B.}~\bibnamefont {Kobrin}}, \bibinfo {author} {\bibfnamefont {T.~O.}\
  \bibnamefont {H{\"{o}}hn}}, \bibinfo {author} {\bibfnamefont {N.~Z.}\
  \bibnamefont {Rui}}, \bibinfo {author} {\bibfnamefont {M.}~\bibnamefont
  {Kamrani}}, \bibinfo {author} {\bibfnamefont {S.}~\bibnamefont {Chatterjee}},
  \bibinfo {author} {\bibfnamefont {S.}~\bibnamefont {Choi}}, \bibinfo {author}
  {\bibfnamefont {M.}~\bibnamefont {Zaletel}}, \bibinfo {author} {\bibfnamefont
  {V.~V.}\ \bibnamefont {Struzhkin}}, \bibinfo {author} {\bibfnamefont {J.~E.}\
  \bibnamefont {Moore}}, \bibinfo {author} {\bibfnamefont {V.~I.}\ \bibnamefont
  {Levitas}}, \bibinfo {author} {\bibfnamefont {R.}~\bibnamefont {Jeanloz}},\
  and\ \bibinfo {author} {\bibfnamefont {N.~Y.}\ \bibnamefont {Yao}},\
  }\bibfield  {title} {\bibinfo {title} {{Imaging stress and magnetism at high
  pressures using a nanoscale quantum sensor}},\ }\href
  {https://doi.org/10.1126/science.aaw4352} {\bibfield  {journal} {\bibinfo
  {journal} {Science}\ }\textbf {\bibinfo {volume} {366}},\ \bibinfo {pages}
  {1349} (\bibinfo {year} {2019})}\BibitemShut {NoStop}%
\bibitem [{\citenamefont {Toyli}\ \emph {et~al.}(2013)\citenamefont {Toyli},
  \citenamefont {de~las Casas}, \citenamefont {Christle}, \citenamefont
  {Dobrovitski},\ and\ \citenamefont {Awschalom}}]{Toyli2013}%
  \BibitemOpen
  \bibfield  {author} {\bibinfo {author} {\bibfnamefont {D.~M.}\ \bibnamefont
  {Toyli}}, \bibinfo {author} {\bibfnamefont {C.~F.}\ \bibnamefont {de~las
  Casas}}, \bibinfo {author} {\bibfnamefont {D.~J.}\ \bibnamefont {Christle}},
  \bibinfo {author} {\bibfnamefont {V.~V.}\ \bibnamefont {Dobrovitski}},\ and\
  \bibinfo {author} {\bibfnamefont {D.~D.}\ \bibnamefont {Awschalom}},\
  }\bibfield  {title} {\bibinfo {title} {{Fluorescence thermometry enhanced by
  the quantum coherence of single spins in diamond}},\ }\href
  {https://doi.org/10.1073/pnas.1306825110} {\bibfield  {journal} {\bibinfo
  {journal} {Proceedings of the National Academy of Sciences}\ }\textbf
  {\bibinfo {volume} {110}},\ \bibinfo {pages} {8417} (\bibinfo {year}
  {2013})}\BibitemShut {NoStop}%
\bibitem [{\citenamefont {Neumann}\ \emph {et~al.}(2013)\citenamefont
  {Neumann}, \citenamefont {Jakobi}, \citenamefont {Dolde}, \citenamefont
  {Burk}, \citenamefont {Reuter}, \citenamefont {Waldherr}, \citenamefont
  {Honert}, \citenamefont {Wolf}, \citenamefont {Brunner}, \citenamefont
  {Shim}, \citenamefont {Suter}, \citenamefont {Sumiya}, \citenamefont
  {Isoya},\ and\ \citenamefont {Wrachtrup}}]{Neumann2013}%
  \BibitemOpen
  \bibfield  {author} {\bibinfo {author} {\bibfnamefont {P.}~\bibnamefont
  {Neumann}}, \bibinfo {author} {\bibfnamefont {I.}~\bibnamefont {Jakobi}},
  \bibinfo {author} {\bibfnamefont {F.}~\bibnamefont {Dolde}}, \bibinfo
  {author} {\bibfnamefont {C.}~\bibnamefont {Burk}}, \bibinfo {author}
  {\bibfnamefont {R.}~\bibnamefont {Reuter}}, \bibinfo {author} {\bibfnamefont
  {G.}~\bibnamefont {Waldherr}}, \bibinfo {author} {\bibfnamefont
  {J.}~\bibnamefont {Honert}}, \bibinfo {author} {\bibfnamefont
  {T.}~\bibnamefont {Wolf}}, \bibinfo {author} {\bibfnamefont {A.}~\bibnamefont
  {Brunner}}, \bibinfo {author} {\bibfnamefont {J.~H.}\ \bibnamefont {Shim}},
  \bibinfo {author} {\bibfnamefont {D.}~\bibnamefont {Suter}}, \bibinfo
  {author} {\bibfnamefont {H.}~\bibnamefont {Sumiya}}, \bibinfo {author}
  {\bibfnamefont {J.}~\bibnamefont {Isoya}},\ and\ \bibinfo {author}
  {\bibfnamefont {J.}~\bibnamefont {Wrachtrup}},\ }\bibfield  {title} {\bibinfo
  {title} {{High-Precision Nanoscale Temperature Sensing Using Single Defects
  in Diamond}},\ }\href {https://doi.org/10.1021/nl401216y} {\bibfield
  {journal} {\bibinfo  {journal} {Nano Letters}\ }\textbf {\bibinfo {volume}
  {13}},\ \bibinfo {pages} {2738} (\bibinfo {year} {2013})}\BibitemShut
  {NoStop}%
\bibitem [{SI()}]{SI}%
  \BibitemOpen
  \href@noop {} {}\bibinfo {note} {Additional details are included in the
  supplemental material.}\BibitemShut {Stop}%
\bibitem [{\citenamefont {Edmonds}\ \emph {et~al.}(2021)\citenamefont
  {Edmonds}, \citenamefont {Hart}, \citenamefont {Turner}, \citenamefont
  {Colard}, \citenamefont {Schloss}, \citenamefont {Olsson}, \citenamefont
  {Trubko}, \citenamefont {Markham}, \citenamefont {Rathmill}, \citenamefont
  {Horne-Smith}, \citenamefont {Lew}, \citenamefont {Manickam}, \citenamefont
  {Bruce}, \citenamefont {Kaup}, \citenamefont {Russo}, \citenamefont
  {DiMario}, \citenamefont {South}, \citenamefont {Hansen}, \citenamefont
  {Twitchen},\ and\ \citenamefont {Walsworth}}]{Edmonds2021}%
  \BibitemOpen
  \bibfield  {author} {\bibinfo {author} {\bibfnamefont {A.~M.}\ \bibnamefont
  {Edmonds}}, \bibinfo {author} {\bibfnamefont {C.~A.}\ \bibnamefont {Hart}},
  \bibinfo {author} {\bibfnamefont {M.~J.}\ \bibnamefont {Turner}}, \bibinfo
  {author} {\bibfnamefont {P.-O.}\ \bibnamefont {Colard}}, \bibinfo {author}
  {\bibfnamefont {J.~M.}\ \bibnamefont {Schloss}}, \bibinfo {author}
  {\bibfnamefont {K.~S.}\ \bibnamefont {Olsson}}, \bibinfo {author}
  {\bibfnamefont {R.}~\bibnamefont {Trubko}}, \bibinfo {author} {\bibfnamefont
  {M.~L.}\ \bibnamefont {Markham}}, \bibinfo {author} {\bibfnamefont
  {A.}~\bibnamefont {Rathmill}}, \bibinfo {author} {\bibfnamefont
  {B.}~\bibnamefont {Horne-Smith}}, \bibinfo {author} {\bibfnamefont
  {W.}~\bibnamefont {Lew}}, \bibinfo {author} {\bibfnamefont {A.}~\bibnamefont
  {Manickam}}, \bibinfo {author} {\bibfnamefont {S.}~\bibnamefont {Bruce}},
  \bibinfo {author} {\bibfnamefont {P.~G.}\ \bibnamefont {Kaup}}, \bibinfo
  {author} {\bibfnamefont {J.~C.}\ \bibnamefont {Russo}}, \bibinfo {author}
  {\bibfnamefont {M.~J.}\ \bibnamefont {DiMario}}, \bibinfo {author}
  {\bibfnamefont {J.~T.}\ \bibnamefont {South}}, \bibinfo {author}
  {\bibfnamefont {J.~T.}\ \bibnamefont {Hansen}}, \bibinfo {author}
  {\bibfnamefont {D.~J.}\ \bibnamefont {Twitchen}},\ and\ \bibinfo {author}
  {\bibfnamefont {R.~L.}\ \bibnamefont {Walsworth}},\ }\bibfield  {title}
  {\bibinfo {title} {Characterisation of {CVD} diamond with high concentrations
  of nitrogen for magnetic-field sensing applications},\ }\href
  {https://doi.org/10.1088/2633-4356/abd88a} {\bibfield  {journal} {\bibinfo
  {journal} {Materials for Quantum Technology}\ }\textbf {\bibinfo {volume}
  {1}},\ \bibinfo {pages} {025001} (\bibinfo {year} {2021})}\BibitemShut
  {NoStop}%
\bibitem [{\citenamefont {Laraoui}\ \emph {et~al.}(2013)\citenamefont
  {Laraoui}, \citenamefont {Dolde}, \citenamefont {Burk}, \citenamefont
  {Reinhard}, \citenamefont {Wrachtrup},\ and\ \citenamefont
  {Meriles}}]{Laraoui2013}%
  \BibitemOpen
  \bibfield  {author} {\bibinfo {author} {\bibfnamefont {A.}~\bibnamefont
  {Laraoui}}, \bibinfo {author} {\bibfnamefont {F.}~\bibnamefont {Dolde}},
  \bibinfo {author} {\bibfnamefont {C.}~\bibnamefont {Burk}}, \bibinfo {author}
  {\bibfnamefont {F.}~\bibnamefont {Reinhard}}, \bibinfo {author}
  {\bibfnamefont {J.}~\bibnamefont {Wrachtrup}},\ and\ \bibinfo {author}
  {\bibfnamefont {C.~A.}\ \bibnamefont {Meriles}},\ }\bibfield  {title}
  {\bibinfo {title} {{High-resolution correlation spectroscopy of 13C spins
  near a nitrogen-vacancy centre in diamond}},\ }\href
  {https://doi.org/10.1038/ncomms2685} {\bibfield  {journal} {\bibinfo
  {journal} {Nature Communications}\ }\textbf {\bibinfo {volume} {4}},\
  \bibinfo {pages} {1651} (\bibinfo {year} {2013})}\BibitemShut {NoStop}%
\bibitem [{\citenamefont {Bucher}\ \emph {et~al.}(2019)\citenamefont {Bucher},
  \citenamefont {Aude~Craik}, \citenamefont {Backlund}, \citenamefont {Turner},
  \citenamefont {Ben~Dor}, \citenamefont {Glenn},\ and\ \citenamefont
  {Walsworth}}]{Bucher2019}%
  \BibitemOpen
  \bibfield  {author} {\bibinfo {author} {\bibfnamefont {D.~B.}\ \bibnamefont
  {Bucher}}, \bibinfo {author} {\bibfnamefont {D.~P.~L.}\ \bibnamefont
  {Aude~Craik}}, \bibinfo {author} {\bibfnamefont {M.~P.}\ \bibnamefont
  {Backlund}}, \bibinfo {author} {\bibfnamefont {M.~J.}\ \bibnamefont
  {Turner}}, \bibinfo {author} {\bibfnamefont {O.}~\bibnamefont {Ben~Dor}},
  \bibinfo {author} {\bibfnamefont {D.~R.}\ \bibnamefont {Glenn}},\ and\
  \bibinfo {author} {\bibfnamefont {R.~L.}\ \bibnamefont {Walsworth}},\
  }\bibfield  {title} {\bibinfo {title} {Quantum diamond spectrometer for
  nanoscale nmr and esr spectroscopy},\ }\href
  {https://doi.org/10.1038/s41596-019-0201-3} {\bibfield  {journal} {\bibinfo
  {journal} {Nature Protocols}\ }\textbf {\bibinfo {volume} {14}},\ \bibinfo
  {pages} {2707} (\bibinfo {year} {2019})}\BibitemShut {NoStop}%
\bibitem [{\citenamefont {Staudacher}\ \emph {et~al.}(2015)\citenamefont
  {Staudacher}, \citenamefont {Raatz}, \citenamefont {Pezzagna}, \citenamefont
  {Meijer}, \citenamefont {Reinhard}, \citenamefont {Meriles},\ and\
  \citenamefont {Wrachtrup}}]{Staudacher2015}%
  \BibitemOpen
  \bibfield  {author} {\bibinfo {author} {\bibfnamefont {T.}~\bibnamefont
  {Staudacher}}, \bibinfo {author} {\bibfnamefont {N.}~\bibnamefont {Raatz}},
  \bibinfo {author} {\bibfnamefont {S.}~\bibnamefont {Pezzagna}}, \bibinfo
  {author} {\bibfnamefont {J.}~\bibnamefont {Meijer}}, \bibinfo {author}
  {\bibfnamefont {F.}~\bibnamefont {Reinhard}}, \bibinfo {author}
  {\bibfnamefont {C.~A.}\ \bibnamefont {Meriles}},\ and\ \bibinfo {author}
  {\bibfnamefont {J.}~\bibnamefont {Wrachtrup}},\ }\bibfield  {title} {\bibinfo
  {title} {{Probing molecular dynamics at the nanoscale via an individual
  paramagnetic centre}},\ }\href {https://doi.org/10.1038/ncomms9527}
  {\bibfield  {journal} {\bibinfo  {journal} {Nature Communications}\ }\textbf
  {\bibinfo {volume} {6}},\ \bibinfo {pages} {8527} (\bibinfo {year}
  {2015})}\BibitemShut {NoStop}%
\bibitem [{\citenamefont {Kong}\ \emph {et~al.}(2015)\citenamefont {Kong},
  \citenamefont {Stark}, \citenamefont {Du}, \citenamefont {McGuinness},\ and\
  \citenamefont {Jelezko}}]{Kong_corr}%
  \BibitemOpen
  \bibfield  {author} {\bibinfo {author} {\bibfnamefont {X.}~\bibnamefont
  {Kong}}, \bibinfo {author} {\bibfnamefont {A.}~\bibnamefont {Stark}},
  \bibinfo {author} {\bibfnamefont {J.}~\bibnamefont {Du}}, \bibinfo {author}
  {\bibfnamefont {L.~P.}\ \bibnamefont {McGuinness}},\ and\ \bibinfo {author}
  {\bibfnamefont {F.}~\bibnamefont {Jelezko}},\ }\bibfield  {title} {\bibinfo
  {title} {Towards chemical structure resolution with nanoscale nuclear
  magnetic resonance spectroscopy},\ }\href
  {https://doi.org/10.1103/PhysRevApplied.4.024004} {\bibfield  {journal}
  {\bibinfo  {journal} {Phys. Rev. Applied}\ }\textbf {\bibinfo {volume} {4}},\
  \bibinfo {pages} {024004} (\bibinfo {year} {2015})}\BibitemShut {NoStop}%
\bibitem [{\citenamefont {Kehayias}\ \emph {et~al.}(2017)\citenamefont
  {Kehayias}, \citenamefont {Jarmola}, \citenamefont {Mosavian}, \citenamefont
  {Fescenko}, \citenamefont {Benito}, \citenamefont {Laraoui}, \citenamefont
  {Smits}, \citenamefont {Bougas}, \citenamefont {Budker}, \citenamefont
  {Neumann}, \citenamefont {Brueck},\ and\ \citenamefont
  {Acosta}}]{Kehayias2017}%
  \BibitemOpen
  \bibfield  {author} {\bibinfo {author} {\bibfnamefont {P.}~\bibnamefont
  {Kehayias}}, \bibinfo {author} {\bibfnamefont {A.}~\bibnamefont {Jarmola}},
  \bibinfo {author} {\bibfnamefont {N.}~\bibnamefont {Mosavian}}, \bibinfo
  {author} {\bibfnamefont {I.}~\bibnamefont {Fescenko}}, \bibinfo {author}
  {\bibfnamefont {F.~M.}\ \bibnamefont {Benito}}, \bibinfo {author}
  {\bibfnamefont {A.}~\bibnamefont {Laraoui}}, \bibinfo {author} {\bibfnamefont
  {J.}~\bibnamefont {Smits}}, \bibinfo {author} {\bibfnamefont
  {L.}~\bibnamefont {Bougas}}, \bibinfo {author} {\bibfnamefont
  {D.}~\bibnamefont {Budker}}, \bibinfo {author} {\bibfnamefont
  {A.}~\bibnamefont {Neumann}}, \bibinfo {author} {\bibfnamefont {S.~R.~J.}\
  \bibnamefont {Brueck}},\ and\ \bibinfo {author} {\bibfnamefont {V.~M.}\
  \bibnamefont {Acosta}},\ }\bibfield  {title} {\bibinfo {title} {{Solution
  nuclear magnetic resonance spectroscopy on a nanostructured diamond chip}},\
  }\href {https://doi.org/10.1038/s41467-017-00266-4} {\bibfield  {journal}
  {\bibinfo  {journal} {Nature Communications}\ }\textbf {\bibinfo {volume}
  {8}},\ \bibinfo {pages} {188} (\bibinfo {year} {2017})}\BibitemShut {NoStop}%
\bibitem [{\citenamefont {Gullion}\ \emph {et~al.}(1990)\citenamefont
  {Gullion}, \citenamefont {Baker},\ and\ \citenamefont
  {Conradi}}]{Gullion_xy8}%
  \BibitemOpen
  \bibfield  {author} {\bibinfo {author} {\bibfnamefont {T.}~\bibnamefont
  {Gullion}}, \bibinfo {author} {\bibfnamefont {D.~B.}\ \bibnamefont {Baker}},\
  and\ \bibinfo {author} {\bibfnamefont {M.~S.}\ \bibnamefont {Conradi}},\
  }\bibfield  {title} {\bibinfo {title} {New, compensated carr-purcell
  sequences},\ }\href
  {https://doi.org/https://doi.org/10.1016/0022-2364(90)90331-3} {\bibfield
  {journal} {\bibinfo  {journal} {Journal of Magnetic Resonance (1969)}\
  }\textbf {\bibinfo {volume} {89}},\ \bibinfo {pages} {479} (\bibinfo {year}
  {1990})}\BibitemShut {NoStop}%
\bibitem [{\citenamefont {Ryan}\ \emph {et~al.}(2010)\citenamefont {Ryan},
  \citenamefont {Hodges},\ and\ \citenamefont {Cory}}]{Ryan_xy8}%
  \BibitemOpen
  \bibfield  {author} {\bibinfo {author} {\bibfnamefont {C.~A.}\ \bibnamefont
  {Ryan}}, \bibinfo {author} {\bibfnamefont {J.~S.}\ \bibnamefont {Hodges}},\
  and\ \bibinfo {author} {\bibfnamefont {D.~G.}\ \bibnamefont {Cory}},\
  }\bibfield  {title} {\bibinfo {title} {Robust decoupling techniques to extend
  quantum coherence in diamond},\ }\href
  {https://doi.org/10.1103/PhysRevLett.105.200402} {\bibfield  {journal}
  {\bibinfo  {journal} {Phys. Rev. Lett.}\ }\textbf {\bibinfo {volume} {105}},\
  \bibinfo {pages} {200402} (\bibinfo {year} {2010})}\BibitemShut {NoStop}%
\bibitem [{\citenamefont {Zhou}\ \emph {et~al.}(2020)\citenamefont {Zhou},
  \citenamefont {Choi}, \citenamefont {Choi}, \citenamefont {Landig},
  \citenamefont {Douglas}, \citenamefont {Isoya}, \citenamefont {Jelezko},
  \citenamefont {Onoda}, \citenamefont {Sumiya}, \citenamefont {Cappellaro},
  \citenamefont {Knowles}, \citenamefont {Park},\ and\ \citenamefont
  {Lukin}}]{Zhou2020}%
  \BibitemOpen
  \bibfield  {author} {\bibinfo {author} {\bibfnamefont {H.}~\bibnamefont
  {Zhou}}, \bibinfo {author} {\bibfnamefont {J.}~\bibnamefont {Choi}}, \bibinfo
  {author} {\bibfnamefont {S.}~\bibnamefont {Choi}}, \bibinfo {author}
  {\bibfnamefont {R.}~\bibnamefont {Landig}}, \bibinfo {author} {\bibfnamefont
  {A.~M.}\ \bibnamefont {Douglas}}, \bibinfo {author} {\bibfnamefont
  {J.}~\bibnamefont {Isoya}}, \bibinfo {author} {\bibfnamefont
  {F.}~\bibnamefont {Jelezko}}, \bibinfo {author} {\bibfnamefont
  {S.}~\bibnamefont {Onoda}}, \bibinfo {author} {\bibfnamefont
  {H.}~\bibnamefont {Sumiya}}, \bibinfo {author} {\bibfnamefont
  {P.}~\bibnamefont {Cappellaro}}, \bibinfo {author} {\bibfnamefont {H.~S.}\
  \bibnamefont {Knowles}}, \bibinfo {author} {\bibfnamefont {H.}~\bibnamefont
  {Park}},\ and\ \bibinfo {author} {\bibfnamefont {M.~D.}\ \bibnamefont
  {Lukin}},\ }\bibfield  {title} {\bibinfo {title} {Quantum metrology with
  strongly interacting spin systems},\ }\href
  {https://doi.org/10.1103/PhysRevX.10.031003} {\bibfield  {journal} {\bibinfo
  {journal} {Phys. Rev. X}\ }\textbf {\bibinfo {volume} {10}},\ \bibinfo
  {pages} {031003} (\bibinfo {year} {2020})}\BibitemShut {NoStop}%
\bibitem [{\citenamefont {Choi}\ \emph {et~al.}(2020)\citenamefont {Choi},
  \citenamefont {Zhou}, \citenamefont {Knowles}, \citenamefont {Landig},
  \citenamefont {Choi},\ and\ \citenamefont {Lukin}}]{Choi2020}%
  \BibitemOpen
  \bibfield  {author} {\bibinfo {author} {\bibfnamefont {J.}~\bibnamefont
  {Choi}}, \bibinfo {author} {\bibfnamefont {H.}~\bibnamefont {Zhou}}, \bibinfo
  {author} {\bibfnamefont {H.~S.}\ \bibnamefont {Knowles}}, \bibinfo {author}
  {\bibfnamefont {R.}~\bibnamefont {Landig}}, \bibinfo {author} {\bibfnamefont
  {S.}~\bibnamefont {Choi}},\ and\ \bibinfo {author} {\bibfnamefont {M.~D.}\
  \bibnamefont {Lukin}},\ }\bibfield  {title} {\bibinfo {title} {Robust dynamic
  hamiltonian engineering of many-body spin systems},\ }\href
  {https://doi.org/10.1103/PhysRevX.10.031002} {\bibfield  {journal} {\bibinfo
  {journal} {Phys. Rev. X}\ }\textbf {\bibinfo {volume} {10}},\ \bibinfo
  {pages} {031002} (\bibinfo {year} {2020})}\BibitemShut {NoStop}%
\bibitem [{\citenamefont {Bauch}\ \emph {et~al.}(2020)\citenamefont {Bauch},
  \citenamefont {Singh}, \citenamefont {Lee}, \citenamefont {Hart},
  \citenamefont {Schloss}, \citenamefont {Turner}, \citenamefont {Barry},
  \citenamefont {Pham}, \citenamefont {Bar-Gill}, \citenamefont {Yelin},\ and\
  \citenamefont {Walsworth}}]{Bauch2020}%
  \BibitemOpen
  \bibfield  {author} {\bibinfo {author} {\bibfnamefont {E.}~\bibnamefont
  {Bauch}}, \bibinfo {author} {\bibfnamefont {S.}~\bibnamefont {Singh}},
  \bibinfo {author} {\bibfnamefont {J.}~\bibnamefont {Lee}}, \bibinfo {author}
  {\bibfnamefont {C.~A.}\ \bibnamefont {Hart}}, \bibinfo {author}
  {\bibfnamefont {J.~M.}\ \bibnamefont {Schloss}}, \bibinfo {author}
  {\bibfnamefont {M.~J.}\ \bibnamefont {Turner}}, \bibinfo {author}
  {\bibfnamefont {J.~F.}\ \bibnamefont {Barry}}, \bibinfo {author}
  {\bibfnamefont {L.~M.}\ \bibnamefont {Pham}}, \bibinfo {author}
  {\bibfnamefont {N.}~\bibnamefont {Bar-Gill}}, \bibinfo {author}
  {\bibfnamefont {S.~F.}\ \bibnamefont {Yelin}},\ and\ \bibinfo {author}
  {\bibfnamefont {R.~L.}\ \bibnamefont {Walsworth}},\ }\bibfield  {title}
  {\bibinfo {title} {Decoherence of ensembles of nitrogen-vacancy centers in
  diamond},\ }\href {https://doi.org/10.1103/PhysRevB.102.134210} {\bibfield
  {journal} {\bibinfo  {journal} {Phys. Rev. B}\ }\textbf {\bibinfo {volume}
  {102}},\ \bibinfo {pages} {134210} (\bibinfo {year} {2020})}\BibitemShut
  {NoStop}%
\bibitem [{\citenamefont {Bucher}\ \emph {et~al.}(2020)\citenamefont {Bucher},
  \citenamefont {Glenn}, \citenamefont {Park}, \citenamefont {Lukin},\ and\
  \citenamefont {Walsworth}}]{Bucher2018}%
  \BibitemOpen
  \bibfield  {author} {\bibinfo {author} {\bibfnamefont {D.~B.}\ \bibnamefont
  {Bucher}}, \bibinfo {author} {\bibfnamefont {D.~R.}\ \bibnamefont {Glenn}},
  \bibinfo {author} {\bibfnamefont {H.}~\bibnamefont {Park}}, \bibinfo {author}
  {\bibfnamefont {M.~D.}\ \bibnamefont {Lukin}},\ and\ \bibinfo {author}
  {\bibfnamefont {R.~L.}\ \bibnamefont {Walsworth}},\ }\bibfield  {title}
  {\bibinfo {title} {Hyperpolarization-enhanced nmr spectroscopy with femtomole
  sensitivity using quantum defects in diamond},\ }\href
  {https://link.aps.org/doi/10.1103/PhysRevX.10.021053} {\bibfield  {journal}
  {\bibinfo  {journal} {Phys. Rev. X}\ }\textbf {\bibinfo {volume} {10}},\
  \bibinfo {pages} {021053} (\bibinfo {year} {2020})}\BibitemShut {NoStop}%
\bibitem [{\citenamefont {Arunkumar}\ \emph {et~al.}(2021)\citenamefont
  {Arunkumar}, \citenamefont {Bucher}, \citenamefont {Turner}, \citenamefont
  {TomHon}, \citenamefont {Glenn}, \citenamefont {Lehmkuhl}, \citenamefont
  {Lukin}, \citenamefont {Park}, \citenamefont {Rosen}, \citenamefont {Theis},\
  and\ \citenamefont {Walsworth}}]{Arunkumar2021}%
  \BibitemOpen
  \bibfield  {author} {\bibinfo {author} {\bibfnamefont {N.}~\bibnamefont
  {Arunkumar}}, \bibinfo {author} {\bibfnamefont {D.~B.}\ \bibnamefont
  {Bucher}}, \bibinfo {author} {\bibfnamefont {M.~J.}\ \bibnamefont {Turner}},
  \bibinfo {author} {\bibfnamefont {P.}~\bibnamefont {TomHon}}, \bibinfo
  {author} {\bibfnamefont {D.}~\bibnamefont {Glenn}}, \bibinfo {author}
  {\bibfnamefont {S.}~\bibnamefont {Lehmkuhl}}, \bibinfo {author}
  {\bibfnamefont {M.~D.}\ \bibnamefont {Lukin}}, \bibinfo {author}
  {\bibfnamefont {H.}~\bibnamefont {Park}}, \bibinfo {author} {\bibfnamefont
  {M.~S.}\ \bibnamefont {Rosen}}, \bibinfo {author} {\bibfnamefont
  {T.}~\bibnamefont {Theis}},\ and\ \bibinfo {author} {\bibfnamefont {R.~L.}\
  \bibnamefont {Walsworth}},\ }\bibfield  {title} {\bibinfo {title}
  {Micron-scale nv-nmr spectroscopy with signal amplification by reversible
  exchange},\ }\href {https://doi.org/10.1103/PRXQuantum.2.010305} {\bibfield
  {journal} {\bibinfo  {journal} {PRX Quantum}\ }\textbf {\bibinfo {volume}
  {2}},\ \bibinfo {pages} {010305} (\bibinfo {year} {2021})}\BibitemShut
  {NoStop}%
\bibitem [{\citenamefont {Castelletto}\ and\ \citenamefont
  {Boretti}(2020)}]{Castelletto2020}%
  \BibitemOpen
  \bibfield  {author} {\bibinfo {author} {\bibfnamefont {S.}~\bibnamefont
  {Castelletto}}\ and\ \bibinfo {author} {\bibfnamefont {A.}~\bibnamefont
  {Boretti}},\ }\bibfield  {title} {\bibinfo {title} {{Silicon carbide color
  centers for quantum applications}},\ }\href
  {https://doi.org/10.1088/2515-7647/ab77a2} {\bibfield  {journal} {\bibinfo
  {journal} {Journal of Physics: Photonics}\ }\textbf {\bibinfo {volume} {2}},\
  \bibinfo {pages} {22001} (\bibinfo {year} {2020})}\BibitemShut {NoStop}%
\bibitem [{\citenamefont {Siyushev}\ \emph {et~al.}(2014)\citenamefont
  {Siyushev}, \citenamefont {Xia}, \citenamefont {Reuter}, \citenamefont
  {Jamali}, \citenamefont {Zhao}, \citenamefont {Yang}, \citenamefont {Duan},
  \citenamefont {Kukharchyk}, \citenamefont {Wieck}, \citenamefont {Kolesov},\
  and\ \citenamefont {Wrachtrup}}]{Siyushev2014}%
  \BibitemOpen
  \bibfield  {author} {\bibinfo {author} {\bibfnamefont {P.}~\bibnamefont
  {Siyushev}}, \bibinfo {author} {\bibfnamefont {K.}~\bibnamefont {Xia}},
  \bibinfo {author} {\bibfnamefont {R.}~\bibnamefont {Reuter}}, \bibinfo
  {author} {\bibfnamefont {M.}~\bibnamefont {Jamali}}, \bibinfo {author}
  {\bibfnamefont {N.}~\bibnamefont {Zhao}}, \bibinfo {author} {\bibfnamefont
  {N.}~\bibnamefont {Yang}}, \bibinfo {author} {\bibfnamefont {C.}~\bibnamefont
  {Duan}}, \bibinfo {author} {\bibfnamefont {N.}~\bibnamefont {Kukharchyk}},
  \bibinfo {author} {\bibfnamefont {A.~D.}\ \bibnamefont {Wieck}}, \bibinfo
  {author} {\bibfnamefont {R.}~\bibnamefont {Kolesov}},\ and\ \bibinfo {author}
  {\bibfnamefont {J.}~\bibnamefont {Wrachtrup}},\ }\bibfield  {title} {\bibinfo
  {title} {{Coherent properties of single rare-earth spin qubits}},\ }\href
  {https://doi.org/10.1038/ncomms4895} {\bibfield  {journal} {\bibinfo
  {journal} {Nature Communications}\ }\textbf {\bibinfo {volume} {5}},\
  \bibinfo {pages} {3895} (\bibinfo {year} {2014})}\BibitemShut {NoStop}%
\bibitem [{\citenamefont {Sajid}\ \emph {et~al.}(2018)\citenamefont {Sajid},
  \citenamefont {Reimers},\ and\ \citenamefont {Ford}}]{Sajid2018}%
  \BibitemOpen
  \bibfield  {author} {\bibinfo {author} {\bibfnamefont {A.}~\bibnamefont
  {Sajid}}, \bibinfo {author} {\bibfnamefont {J.~R.}\ \bibnamefont {Reimers}},\
  and\ \bibinfo {author} {\bibfnamefont {M.~J.}\ \bibnamefont {Ford}},\
  }\bibfield  {title} {\bibinfo {title} {{Defect states in hexagonal boron
  nitride: Assignments of observed properties and prediction of properties
  relevant to quantum computation}},\ }\href
  {https://doi.org/10.1103/PhysRevB.97.064101} {\bibfield  {journal} {\bibinfo
  {journal} {Phys. Rev. B}\ }\textbf {\bibinfo {volume} {97}},\ \bibinfo
  {pages} {64101} (\bibinfo {year} {2018})}\BibitemShut {NoStop}%
\bibitem [{\citenamefont {Udvarhelyi}\ \emph {et~al.}(2017)\citenamefont
  {Udvarhelyi}, \citenamefont {Thiering}, \citenamefont {Londero},\ and\
  \citenamefont {Gali}}]{Udvarhelyi2017}%
  \BibitemOpen
  \bibfield  {author} {\bibinfo {author} {\bibfnamefont {P.}~\bibnamefont
  {Udvarhelyi}}, \bibinfo {author} {\bibfnamefont {G.~m.~H.}\ \bibnamefont
  {Thiering}}, \bibinfo {author} {\bibfnamefont {E.}~\bibnamefont {Londero}},\
  and\ \bibinfo {author} {\bibfnamefont {A.}~\bibnamefont {Gali}},\ }\bibfield
  {title} {\bibinfo {title} {Ab initio theory of the
  ${\mathrm{n}}_{2}\mathrm{V}$ defect in diamond for quantum memory
  implementation},\ }\href {https://doi.org/10.1103/PhysRevB.96.155211}
  {\bibfield  {journal} {\bibinfo  {journal} {Phys. Rev. B}\ }\textbf {\bibinfo
  {volume} {96}},\ \bibinfo {pages} {155211} (\bibinfo {year}
  {2017})}\BibitemShut {NoStop}%
\bibitem [{\citenamefont {Ruskuc}\ \emph {et~al.}(2022)\citenamefont {Ruskuc},
  \citenamefont {Wu}, \citenamefont {Rochman}, \citenamefont {Choi},\ and\
  \citenamefont {Faraon}}]{Ruskuc2022}%
  \BibitemOpen
  \bibfield  {author} {\bibinfo {author} {\bibfnamefont {A.}~\bibnamefont
  {Ruskuc}}, \bibinfo {author} {\bibfnamefont {C.-J.}\ \bibnamefont {Wu}},
  \bibinfo {author} {\bibfnamefont {J.}~\bibnamefont {Rochman}}, \bibinfo
  {author} {\bibfnamefont {J.}~\bibnamefont {Choi}},\ and\ \bibinfo {author}
  {\bibfnamefont {A.}~\bibnamefont {Faraon}},\ }\bibfield  {title} {\bibinfo
  {title} {{Nuclear spin-wave quantum register for a solid-state qubit}},\
  }\href {https://doi.org/10.1038/s41586-021-04293-6} {\bibfield  {journal}
  {\bibinfo  {journal} {Nature}\ }\textbf {\bibinfo {volume} {602}},\ \bibinfo
  {pages} {408} (\bibinfo {year} {2022})}\BibitemShut {NoStop}%
\bibitem [{\citenamefont {Degen}\ \emph {et~al.}(2017)\citenamefont {Degen},
  \citenamefont {Reinhard},\ and\ \citenamefont {Cappellaro}}]{degen_review}%
  \BibitemOpen
  \bibfield  {author} {\bibinfo {author} {\bibfnamefont {C.~L.}\ \bibnamefont
  {Degen}}, \bibinfo {author} {\bibfnamefont {F.}~\bibnamefont {Reinhard}},\
  and\ \bibinfo {author} {\bibfnamefont {P.}~\bibnamefont {Cappellaro}},\
  }\bibfield  {title} {\bibinfo {title} {Quantum sensing},\ }\href
  {https://doi.org/10.1103/RevModPhys.89.035002} {\bibfield  {journal}
  {\bibinfo  {journal} {Rev. Mod. Phys.}\ }\textbf {\bibinfo {volume} {89}},\
  \bibinfo {pages} {035002} (\bibinfo {year} {2017})}\BibitemShut {NoStop}%
\end{thebibliography}%


\begin{thebibliography}{4}%
\makeatletter
\providecommand \@ifxundefined [1]{%
 \@ifx{#1\undefined}
}%
\providecommand \@ifnum [1]{%
 \ifnum #1\expandafter \@firstoftwo
 \else \expandafter \@secondoftwo
 \fi
}%
\providecommand \@ifx [1]{%
 \ifx #1\expandafter \@firstoftwo
 \else \expandafter \@secondoftwo
 \fi
}%
\providecommand \natexlab [1]{#1}%
\providecommand \enquote  [1]{``#1''}%
\providecommand \bibnamefont  [1]{#1}%
\providecommand \bibfnamefont [1]{#1}%
\providecommand \citenamefont [1]{#1}%
\providecommand \href@noop [0]{\@secondoftwo}%
\providecommand \href [0]{\begingroup \@sanitize@url \@href}%
\providecommand \@href[1]{\@@startlink{#1}\@@href}%
\providecommand \@@href[1]{\endgroup#1\@@endlink}%
\providecommand \@sanitize@url [0]{\catcode `\\12\catcode `\$12\catcode
  `\&12\catcode `\#12\catcode `\^12\catcode `\_12\catcode `\%12\relax}%
\providecommand \@@startlink[1]{}%
\providecommand \@@endlink[0]{}%
\providecommand \url  [0]{\begingroup\@sanitize@url \@url }%
\providecommand \@url [1]{\endgroup\@href {#1}{\urlprefix }}%
\providecommand \urlprefix  [0]{URL }%
\providecommand \Eprint [0]{\href }%
\providecommand \doibase [0]{https://doi.org/}%
\providecommand \selectlanguage [0]{\@gobble}%
\providecommand \bibinfo  [0]{\@secondoftwo}%
\providecommand \bibfield  [0]{\@secondoftwo}%
\providecommand \translation [1]{[#1]}%
\providecommand \BibitemOpen [0]{}%
\providecommand \bibitemStop [0]{}%
\providecommand \bibitemNoStop [0]{.\EOS\space}%
\providecommand \EOS [0]{\spacefactor3000\relax}%
\providecommand \BibitemShut  [1]{\csname bibitem#1\endcsname}%
\let\auto@bib@innerbib\@empty
%</preamble>
\bibitem [{\citenamefont {Glenn}\ \emph {et~al.}(2018)\citenamefont {Glenn},
  \citenamefont {Bucher}, \citenamefont {Lee}, \citenamefont {Lukin},
  \citenamefont {Park},\ and\ \citenamefont {Walsworth}}]{Glenn2018}%
  \BibitemOpen
  \bibfield  {author} {\bibinfo {author} {\bibfnamefont {D.~R.}\ \bibnamefont
  {Glenn}}, \bibinfo {author} {\bibfnamefont {D.~B.}\ \bibnamefont {Bucher}},
  \bibinfo {author} {\bibfnamefont {J.}~\bibnamefont {Lee}}, \bibinfo {author}
  {\bibfnamefont {M.~D.}\ \bibnamefont {Lukin}}, \bibinfo {author}
  {\bibfnamefont {H.}~\bibnamefont {Park}},\ and\ \bibinfo {author}
  {\bibfnamefont {R.~L.}\ \bibnamefont {Walsworth}},\ }\bibfield  {title}
  {\bibinfo {title} {High-resolution magnetic resonance spectroscopy using a
  solid-state spin sensor},\ }\href {https://doi.org/10.1038/nature25781}
  {\bibfield  {journal} {\bibinfo  {journal} {Nature}\ }\textbf {\bibinfo
  {volume} {555}},\ \bibinfo {pages} {351} (\bibinfo {year}
  {2018})}\BibitemShut {NoStop}%
\bibitem [{\citenamefont {Zhou}\ \emph {et~al.}(2020)\citenamefont {Zhou},
  \citenamefont {Choi}, \citenamefont {Choi}, \citenamefont {Landig},
  \citenamefont {Douglas}, \citenamefont {Isoya}, \citenamefont {Jelezko},
  \citenamefont {Onoda}, \citenamefont {Sumiya}, \citenamefont {Cappellaro},
  \citenamefont {Knowles}, \citenamefont {Park},\ and\ \citenamefont
  {Lukin}}]{Zhou2020}%
  \BibitemOpen
  \bibfield  {author} {\bibinfo {author} {\bibfnamefont {H.}~\bibnamefont
  {Zhou}}, \bibinfo {author} {\bibfnamefont {J.}~\bibnamefont {Choi}}, \bibinfo
  {author} {\bibfnamefont {S.}~\bibnamefont {Choi}}, \bibinfo {author}
  {\bibfnamefont {R.}~\bibnamefont {Landig}}, \bibinfo {author} {\bibfnamefont
  {A.~M.}\ \bibnamefont {Douglas}}, \bibinfo {author} {\bibfnamefont
  {J.}~\bibnamefont {Isoya}}, \bibinfo {author} {\bibfnamefont
  {F.}~\bibnamefont {Jelezko}}, \bibinfo {author} {\bibfnamefont
  {S.}~\bibnamefont {Onoda}}, \bibinfo {author} {\bibfnamefont
  {H.}~\bibnamefont {Sumiya}}, \bibinfo {author} {\bibfnamefont
  {P.}~\bibnamefont {Cappellaro}}, \bibinfo {author} {\bibfnamefont {H.~S.}\
  \bibnamefont {Knowles}}, \bibinfo {author} {\bibfnamefont {H.}~\bibnamefont
  {Park}},\ and\ \bibinfo {author} {\bibfnamefont {M.~D.}\ \bibnamefont
  {Lukin}},\ }\bibfield  {title} {\bibinfo {title} {Quantum metrology with
  strongly interacting spin systems},\ }\href
  {https://doi.org/10.1103/PhysRevX.10.031003} {\bibfield  {journal} {\bibinfo
  {journal} {Phys. Rev. X}\ }\textbf {\bibinfo {volume} {10}},\ \bibinfo
  {pages} {031003} (\bibinfo {year} {2020})}\BibitemShut {NoStop}%
\bibitem [{\citenamefont {Choi}\ \emph {et~al.}(2020)\citenamefont {Choi},
  \citenamefont {Zhou}, \citenamefont {Knowles}, \citenamefont {Landig},
  \citenamefont {Choi},\ and\ \citenamefont {Lukin}}]{Choi2020}%
  \BibitemOpen
  \bibfield  {author} {\bibinfo {author} {\bibfnamefont {J.}~\bibnamefont
  {Choi}}, \bibinfo {author} {\bibfnamefont {H.}~\bibnamefont {Zhou}}, \bibinfo
  {author} {\bibfnamefont {H.~S.}\ \bibnamefont {Knowles}}, \bibinfo {author}
  {\bibfnamefont {R.}~\bibnamefont {Landig}}, \bibinfo {author} {\bibfnamefont
  {S.}~\bibnamefont {Choi}},\ and\ \bibinfo {author} {\bibfnamefont {M.~D.}\
  \bibnamefont {Lukin}},\ }\bibfield  {title} {\bibinfo {title} {Robust dynamic
  hamiltonian engineering of many-body spin systems},\ }\href
  {https://doi.org/10.1103/PhysRevX.10.031002} {\bibfield  {journal} {\bibinfo
  {journal} {Phys. Rev. X}\ }\textbf {\bibinfo {volume} {10}},\ \bibinfo
  {pages} {031002} (\bibinfo {year} {2020})}\BibitemShut {NoStop}%
\bibitem [{\citenamefont {Jiang}\ \emph {et~al.}(2009)\citenamefont {Jiang},
  \citenamefont {Hodges}, \citenamefont {Maze}, \citenamefont {Maurer},
  \citenamefont {Taylor}, \citenamefont {Cory}, \citenamefont {Hemmer},
  \citenamefont {Walsworth}, \citenamefont {Yacoby}, \citenamefont {Zibrov},\
  and\ \citenamefont {Lukin}}]{Jiang2009}%
  \BibitemOpen
  \bibfield  {author} {\bibinfo {author} {\bibfnamefont {L.}~\bibnamefont
  {Jiang}}, \bibinfo {author} {\bibfnamefont {J.~S.}\ \bibnamefont {Hodges}},
  \bibinfo {author} {\bibfnamefont {J.~R.}\ \bibnamefont {Maze}}, \bibinfo
  {author} {\bibfnamefont {P.}~\bibnamefont {Maurer}}, \bibinfo {author}
  {\bibfnamefont {J.~M.}\ \bibnamefont {Taylor}}, \bibinfo {author}
  {\bibfnamefont {D.~G.}\ \bibnamefont {Cory}}, \bibinfo {author}
  {\bibfnamefont {P.~R.}\ \bibnamefont {Hemmer}}, \bibinfo {author}
  {\bibfnamefont {R.~L.}\ \bibnamefont {Walsworth}}, \bibinfo {author}
  {\bibfnamefont {A.}~\bibnamefont {Yacoby}}, \bibinfo {author} {\bibfnamefont
  {A.~S.}\ \bibnamefont {Zibrov}},\ and\ \bibinfo {author} {\bibfnamefont
  {M.~D.}\ \bibnamefont {Lukin}},\ }\bibfield  {title} {\bibinfo {title}
  {Repetitive readout of a single electronic spin via quantum logic with
  nuclear spin ancillae},\ }\href {https://doi.org/10.1126/science.1176496}
  {\bibfield  {journal} {\bibinfo  {journal} {Science}\ }\textbf {\bibinfo
  {volume} {326}},\ \bibinfo {pages} {267} (\bibinfo {year}
  {2009})}\BibitemShut {NoStop}%
\end{thebibliography}%

\end{document}

\title{Supplement: Quantum Logic Enhanced Sensing in Solid State Spin Ensembles}

\date{\today}

\author{Nithya Arunkumar}
\affiliation{\huPhys}
\affiliation{\huSEAS}
\affiliation{\umdQTC}

\author{Kevin S. Olsson}
\affiliation{\umdQTC}
\affiliation{\umdECE}
\affiliation{\ICpostdoc}

\author{Jner Tzern Oon}
\affiliation{\umdQTC}
\affiliation{\umdPhys}

\author{Connor Hart}
\affiliation{\umdQTC}
\affiliation{\umdECE}

\author{Dominik B. Bucher}
\affiliation{\huPhys}
\affiliation{\tum}

\author{David Glenn}
\affiliation{\huPhys}

\author{Mikhail D. Lukin}
\affiliation{\huPhys}

\author{Hongkun Park}
\affiliation{\huPhys}
\affiliation{\huChem}

\author{Donhee Ham }
\affiliation{\huSEAS}

\author{Ronald L. Walsworth}
\thanks{walsworth@umd.edu}
\affiliation{\huPhys}
\affiliation{\umdQTC}
\affiliation{\umdECE}
\affiliation{\umdPhys}

\maketitle

\tableofcontents

\section{Experimental Methods}
\label{sec:methods}

The ensemble NV sensor is a ($2\times2\times0.5$)$\,$mm$^3$ high purity diamond chip with 13 \textmu m thick, nitrogen-doped layer ([N]$\,\approx\,$14\,ppm)  grown using a high-purity, chemical vapour deposition process (Element Six, Ltd). After irradiation and annealing, the nitrogen-doped layer contains [NV]$\,\approx\,$2.3\,ppm. The NV ensemble dephasing time $T_2^*$, measured using Ramsey spectroscopy, is $T_2^* \approx$ 600\,ns. The NV ensemble $T_2$ decoherence time, measured using a Hahn-echo sequence, is 14.5 \textmu s. The top face of the diamond is cut perpendicular to the [100] crystal axis and the lateral faces are perpendicular to [110]. All four edges of the top face were then polished at 45 degrees (Delaware Diamond Knives), with a top face area of 1 mm $\times$ 1 mm. The [111] axis of the NV sensor is aligned parallel to the bias magnetic field. A variable bias magnetic field (0\,G to 4000\,G) is generated by a feedback-stabilized electromagnet. Details about the magnetic bias field stabilization is described in the Methods section of~\cite{Glenn2018}.

The 130\,mW optical beam ($\lambda$ = 532\,nm), generated by a solid-state laser (Coherent Verdi G7), is focused down to a spot size of about 15 \textmu m and pulsed using an acousto-optic modulator (Gooch \& Housego, 3250-220). For the experiments in this work, the duration of each optical pulse is 3 \textmu s. The NV spin-state-dependent fluorescence is read out after 1 \textmu s, followed by additional optical re-initialization of the NV electronic spin for 2 \textmu s. The NV fluorescence signal is collected by a liquid light guide (Thorlabs LLG5-8H) and delivered to a photodetector (Thorlabs PDB210A). An arbitrary waveform generator (Tektronix AWG7122C) generates microwave (MW) pulses with a temporal resolution of 12 Giga-samples per second (83 ps). The pulses are amplified using a  microwave amplifier (Minicircuits ZHL-25W-63+, RF Lambda RFLUPA0618GD). The radiofrequency (RF) pulses (1-3\,MHz), used to apply the cNOT operation on the nuclear spin, conditioned on the electronic spin state, are generated by a function generator (Rigol DG1032). Both the MW and the RF control pulses are combined using a power splitter (Minicircuits ZN2PD2-14W-S+). The combined signal is delivered to the NV diamond, to drive the NV electronic spins and the $^{15}$N nuclear spins, through a single loop shorted coil of 1 mm diameter. A synthetic AC magnetic signal (1 MHz) is generated by applying an AC voltage (Rigol DG1022) to a home-built multi loop test coil (4 loops, 10 mm diameter).

\section{AC Magnetometry}
\label{sec:acmag}

\begin{figure}
    \centering
    \includegraphics[width=\textwidth]{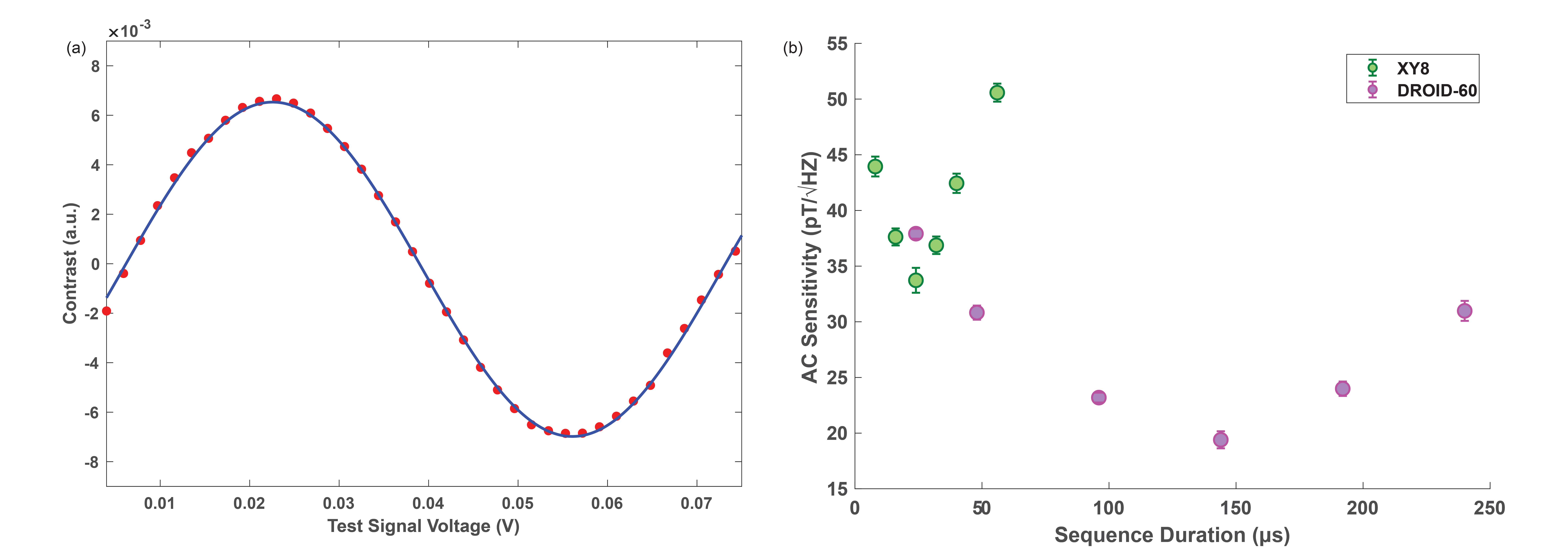}
	\caption{
		\textbf{AC Magnetometry with XY8 and DROID-60.}
		\textbf{(a)} Test signal used in calibration of sensitivity measurements.
		\textbf{(b)} Measured sensitivities for XY8 and DROID-60 decoupling sequences as sequence duration (number of sequence repetitions) is extended.
		}
	\label{fig:figS1}
\end{figure}

We characterized the optimal sequence duration for XY8 and DROID-60 sequences by measuring the AC sensitivity for various sequence durations. The AC signal is generated by applying a voltage to a test coil. The AC magnetic
field from the test coil is first calibrated by measuring the signal amplitude (i.e., the NV contrast) as a function of applied voltage. The NV contrast undergoes periodic oscillations with increasing test signal voltage (red dots in the Fig.~\ref{fig:figS1}(a)). A single oscillation corresponds to 2$\pi$ phase accumulation by the NVs during the sensing sequence. The magnetic signal amplitude that produces 2$\pi$ phase accumulation is given by

\begin{equation}
\text{B}_{AC}(2\pi) =  2*\dfrac{\hbar\pi^2f_0}{g\mu_BN},
\label{eqn:equ1}
\end{equation}

where $g$ = 2 is the Landé g-factor, $\mu_B$ is the Bohr magneton, $f_0$ is the frequency of the sensing sequence (1/2$\tau$), and $N$ is the number of $\pi$ pulses. For DROID-60 at $f_0$ = 1 MHz, $\text{B}_{AC}(2\pi)$  = 0.3891 $\mu$T. From the sinusoidal fit to the data (blue line in Fig.~\ref{fig:figS1}(a)), the 2$\pi$ phase accumulation occurs at 0.0670 V. Comparing these two, we get the calibration constant $\text{B}_{AC}(2\pi)$ / V = 5.806 $\mu$T/V.

The AC sensitivity $\eta$ can be determined by,

\begin{equation}
\eta =  \dfrac{\sigma^{1s}}{|dS/d\text{B}_{AC}|}
\label{eqn:equ2}
\end{equation}
where $\sigma^{1s}$ is the uncertainty of the contrast for 1 second averaging and $|dS/d\text{B}_{AC}|$ is the gradient of contrast with respect to $\text{B}_{AC}$. Both these parameters are calculated at the zero-crossing. $\sigma^{1s}$ is measured from the standard error at the zero-crossing, by acquiring the data for 1 second. $|dS/d\text{B}_{AC}|$ is determined from the slope of the curve near the zero crossing.

The AC sensitivity measurement for XY8 (green circles) and DROID-60 (purple circles) is shown in Fig.~\ref{fig:figS1}(b). The sensing duration is on the horizontal axis, which is determined by the number of repetitions of the sequence. The best sensitivity is achieved for DROID-60:6 and XY8:6 for a sensing duration of 144\,$\mu$s and 24\,$\mu$s respectively. For longer sensing durations, the DROID-60 sequence outperforms the XY8 sequence.

\begin{figure}[h]
    \centering
    \includegraphics[width=1.75 in]{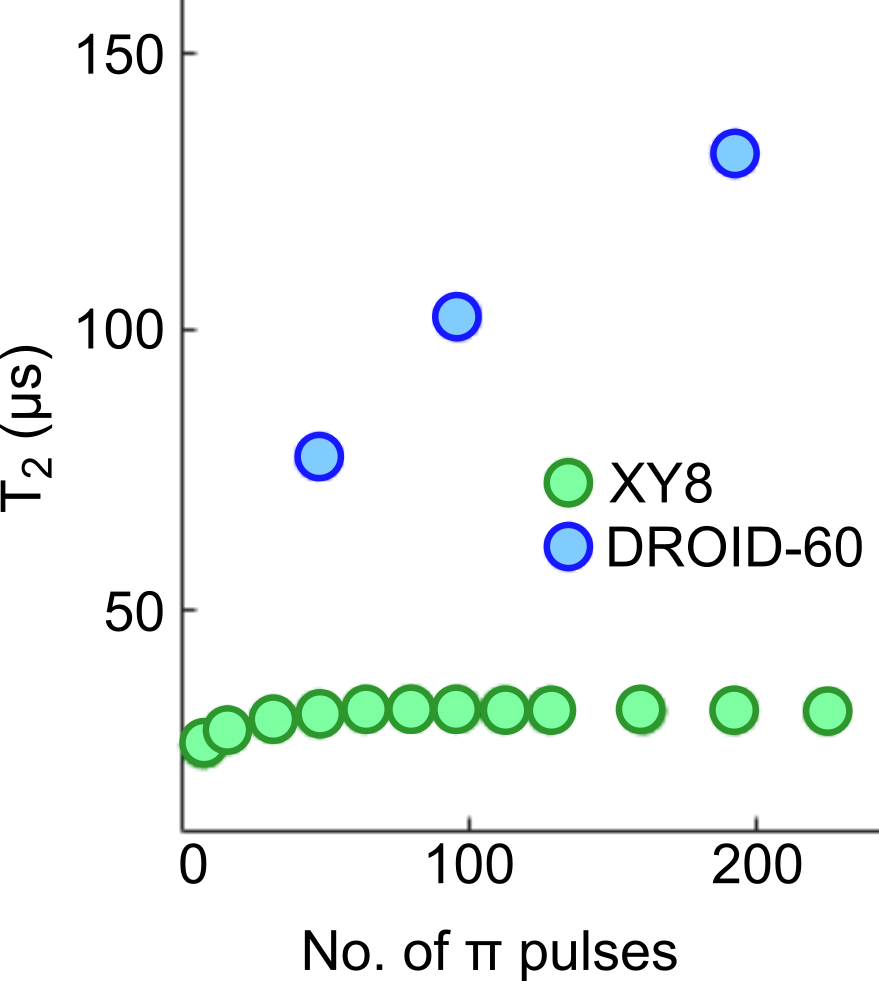}
	\caption{
		\textbf{NV Electronic Spin Coherence Time $T_\text{2}$.} Measured NV $T_\text{2}$ under XY8 and DROID-60 dynamical decoupling as a function of the number of applied $\pi$-pulses. The XY8 $T_\text{2}$ saturation is attributed to like-spin, NV-NV interactions.
		}
	\label{fig:figS2}
\end{figure}

The improved AC sensitivity achieved with the DROID-60 is attributed to an extended NV electronic spin coherence time $T_2$ under DROID decoupling, as shown in Fig.~\ref{fig:figS2} for both sequences. Under XY8 decoupling, the measured NV $T_2$ saturates at about 28 \textmu s due to like-spin, NV-NV interactions (i.e., instantaneous diffusion) which are not decoupled by the XY8 sequence. Meanwhile, as previously demonstrated~\cite{Zhou2020,Choi2020}, the DROID-60 sequence does decouple such interactions, enabling increased $T_2$ with the application of additional decoupling cycles.

\section{Signal-to-Noise Ratio with Quantum Logic Enhanced Readout}
\label{sec:calc}

In comparing $N$ conventional readouts, the amplitude of each readout $A_n$ corresponds to a measurement of the same spin population. Thus, the signal-to-noise ratio (SNR) for the mean signal of N measurements is simply the average signal divided by the root mean squared error

$$
\text{SNR}(N)_{con} = \dfrac{\sum_{n=1}^N A_n}{\sqrt{\sum_{n=1}^N \sigma_n^2}}
$$
where $\sigma_n$ is the error (uncertainty) of the $n$-th measurement. Now with the quantum logic enhanced (QLE) readout, $A_n$ corresponds to the decaying spin population of the memory spins, due to their finite lifetime. However, $\sigma_n$ will remain approximately equivalent. This effect can be accounted for by properly weighting the amplitudes and uncertainties of each measurement when calculating the SNR

$$
\text{SNR}(N)_{\text{QLE}} = \dfrac{\sum_{n=1}^N w_n A_n}{\sqrt{\sum_{n=1}^N w_n^2 \sigma_n^2}}
$$
where $w_n$ is the weight. Rewriting the numerator and invoking the Cauchy-Schwarz inequality yields
$$
\left( \sum w_n\sigma_n \frac{A_n}{\sigma_n} \right)^2 \leq
    \left( \sum w_n^2 \sigma_n^2 \right) \left( \sum \frac{A_n^2}{\sigma_n^2}\right).
$$
The equality is fulfilled by when $w_n = A_n / \sigma_n^2$, leading the optimal SNR to be \cite{Jiang2009}
$$
\text{Optimal } \text{SNR}(N)_{\text{QLE}} = \sqrt{ \sum_{n=1}^N \dfrac{ A_n^2}{\sigma_n^2}}.
$$
The weighting factor $A_n / \sigma_n^2$ agrees with the intuition that measurements with lower signal amplitude should be weighted less when calculating SNR$(N)_{\text{QLE}}$ than higher signal amplitudes.

\section{Laser Power Vs Nuclear spin T1}
\label{sec:Laser}

\begin{figure}[h]
    \centering
    \includegraphics[width=4in]{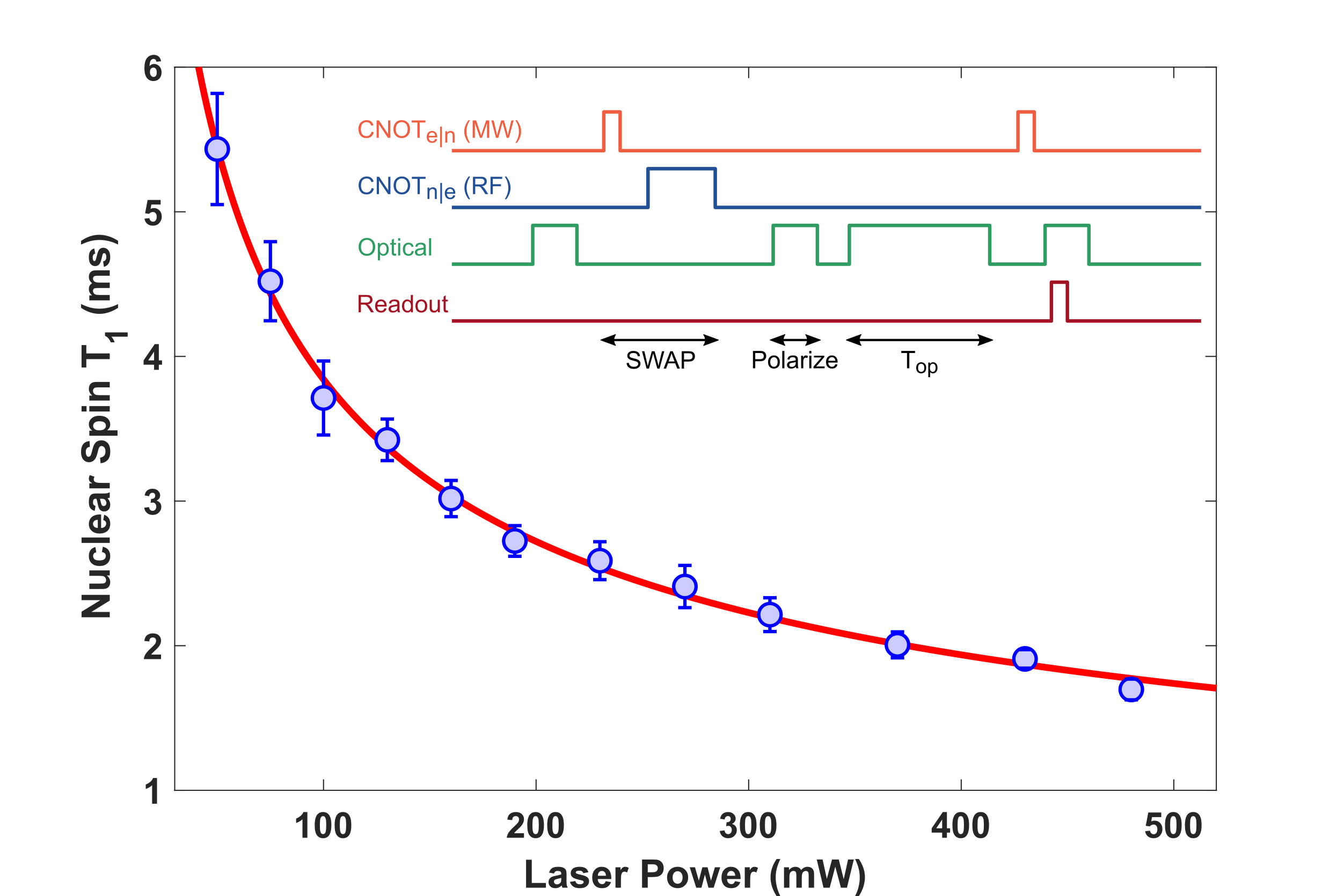}
	\caption{
		\textbf{Laser power versus $^{15}N$ $T_1$}. The pulse sequence used to measure the nuclear $T_1$ at each laser power (blue circles) is shown as an inset. A fixed illumination spot of 15 \textmu m in diameter was used.
		}
	\label{fig:figS3}
\end{figure}

We characterized the nuclear spin life time $T_1$ under optical illumination in the experimental setup used throughout this work. After the nuclear spin swap at a magnetic field of 3700\,G, NV fluorescence contrast signal is measured as a function of optical (AOM) pulse duration $T_\text{op}$, which results in a decay curve. Fitting this curve with a stretched exponential function, yields the $T_1$ of the nuclear spin under optical illumination. The nuclear spin $T_1$ (blue dots) is measured for various laser powers and shown in Fig.~\ref{fig:figS3}. The solid red curve is a power function model of the form $a|x|^{-b}+c$ with fit parameters a = 4.003e+04, b = 0.5154, and c = 111.

While Fig.~\ref{fig:figS3} suggests it would be ideal to work at lower laser power (for a fixed illumination area), since the nuclear spin lifetime is longer at lower laser powers, the effect of increased re-polarization optical pulse durations at lower laser power must also be considered. For QLE sensing, the performance of the technique is maximized by accommodating more readout pulses. This can be achieved by reducing the duration of each readout/re-initialization pulse. To balance these two factors, we choose to work at a laser power of 130 mW where the $\sim$75$\%$ NV spins are re-polarized within 3 \textmu s.

\bibliography{QLE_bibfile}